\documentclass[peerreview, 11pt]{IEEEtran}
\usepackage{graphics}
\usepackage{epsfig}
\renewcommand{\QED}{{\QEDopen}}

\renewcommand{\baselinestretch}{1.48}
\begin{document}

\newtheorem{lemma}{Lemma}
\newtheorem{corollary}{Corollary}
\newtheorem{theorem}{Theorem}
\newtheorem{definition}{Definition}
\newtheorem{algorithm}{Algorithm}
\newtheorem{remark}{Remark}
\newcommand{\bre}{\begin{equation}}
\newcommand{\ere}{\end{equation}}
\newcommand{\be}{{\bf {e}}}
\newcommand{\ee}\]
\newcommand{\bra}{\begin{eqnarray}}
\newcommand{\era}{\end{eqnarray}}
\newcommand{\bfg}{\begin{figure}[hbtp]}
\newcommand{\efg}{\end{figure}}
\newcommand{\bver}{\begin{verbatim}}
\newcommand{\ever}{\end{verbatim}}
\newcommand{\bit}{\begin{itemize}}
\newcommand{\eit}{\end{itemize}}
\newcommand{\ben}{\begin{enumerate}}
\newcommand{\een}{\end{enumerate}}
\newcommand{\ett}{\mbox{$\eta$} }
\newcommand{\coeff}[1]{\lfloor #1\rfloor}
\newcommand{\ceil}[1]{\lceil #1\rceil}
\newcommand{\floor}[1]{[ #1 ]}
\newcommand{\bfloor}[1]{\left[ #1 \right]}
\newcommand{\dgr}[1]{\mbox{$#1^{\circ}$}}
\newcommand{\cu}{\mbox{cosmic $\mu$ }}
\newcommand{\csa}[2]{\mbox{$\cos^2(#1 - #2)$}}
\newcommand{\csb}[2]{\mbox{$\cos2(#1 - #2)$}}
\newcommand{\balpha}{\mbox{\boldmath $\alpha$}}
\newcommand{\bbeta}{\mbox{\boldmath $\beta$}}
\newcommand{\blambda}{\mbox{\boldmath $\lambda$}}
\newcommand{\bkappa}{\mbox{\boldmath $\kappa$}}
\newcommand{\bXi}{\mbox{\boldmath $\Xi$}}
\newcommand{\brho}{\mbox{\boldmath $\rho$}}
\newcommand{\bchi}{\mbox{\bf x}}
\newcommand{\bal}{\mbox{\boldmath $\alpha$}_0}
\newcommand{\Exp}{\mbox{E}}
\newcommand{\given}{\: | \:}
\newcommand{\Ker}{\mbox{Ker}\,}
\newcommand{\tildepss}{\tilde \epsilon_{s}}
\newcommand{\tildepssl}{\tilde \epsilon_{s_l}}
\newcommand{\epss}{\epsilon_s}
\newcommand{\epssl}{\epsilon_{s_l}}
\newcommand{\bphi}{{\mathbf \Phi}}
\newcommand{\bepsilon}{\mbox{\boldmath $\epsilon$}}
\newcommand{\bOmega}{{\mathbf \Omega}}
\newcommand{\btOmega}{\tilde{\mathbf \Omega}}
\newcommand{\bSigma}{{\mathbf \Sigma}}

\newcommand{\bsigma}{\mbox{\boldmath $\sigma$}}
\newcommand{\btsigma}{\mbox{\boldmath $\tilde\sigma$}}

\newcommand{\we}{\overrightarrow{e}}
\newcommand{\bA}{{\bf A}}
\newcommand{\bB}{{\bf B}}
\newcommand{\bC}{{\bf C}}
\newcommand{\bI}{{\bf I}}
\newcommand{\bg}{{\bf{g}}}
\newcommand{\bG}{{\bf{G}}}
\newcommand{\bE}{{\bf E}}
\newcommand{\bF}{{\bf F}}
\newcommand{\bm}{{\bf m}}
\newcommand{\bk}{{\bf k}}
\newcommand{\bof}{{\bf f}}
\newcommand{\rmf}{{\rm f}}
\newcommand{\norm}[1]{\|#1\|}
\newcommand{\oQ}{\overline Q}
\newcommand{\tQ}{\tilde Q}
\newcommand{\tD}{\tilde D}
\newcommand{\oD}{\bar D}
\newcommand{\oV}{\bar V}
\newcommand{\tP}{\tilde P}
\newcommand{\tN}{\tilde N}
\newcommand{\tA}{\tilde A}
\newcommand{\tM}{\tilde M}
\newcommand{\tX}{\tilde X}
\newcommand{\tY}{\tilde Y}
\newcommand{\tW}{\tilde W}
\newcommand{\tm}{\tilde m}
\newcommand{\tc}{\tilde c}
\newcommand{\tv}{\tilde v}
\newcommand{\tH}{\tilde H}
\newcommand{\btW}{\tilde {\bf W}}
\newcommand{\btA}{\tilde {\bf A}}
\newcommand{\btB}{\tilde {\bf B}}
\newcommand{\btV}{\tilde {\bf V}}
\newcommand{\btH}{\tilde {\bf H}}
\newcommand{\btZ}{\tilde {\bf Z}}
\newcommand{\btY}{\tilde {\bf Y}}
\newcommand{\btM}{\tilde {\bf M}}
\newcommand{\btX}{\tilde {\bf X}}
\newcommand{\btx}{\tilde {\bf x}}
\newcommand{\btm}{\tilde {\bf m}}
\newcommand{\bFDPC}{\bF_{DPC}}
\newcommand{\tdet}{\tilde {\textrm{det}}}
\newcommand{\btT}{\tilde {\bf T}}
\newcommand{\btS}{\tilde {\bf S}}
\newcommand{\btn}{\tilde {\bf n}}
\newcommand{\btv}{\tilde {\bf v}}
\newcommand{\bty}{\tilde {\bf y}}
\newcommand{\btp}{\tilde {\bf p}}
\newcommand{\bta}{\tilde {\bf a}}
\newcommand{\PrEras}{\hat{\Pr}}
\newcommand{\ErasEpsilon}{\hat{\epsilon}}
\newcommand{\ErasCY}{\hat{\cY}}
\newcommand{\tab}{\ \ \ \ }
\newcommand{\bigtab}{\tab \tab \tab}
\newcommand{\bc}{{\bf c}}
\newcommand{\type}{{\mathrm type}}
\newcommand{\QEC}{{\mathrm{QEC}}}
\newcommand{\scnd}{{\mathrm {scnd}}}
\newcommand{\APP}{{\mathrm {APP}}}
\newcommand{\LLR}{{\mathrm {LLR}}}
\newcommand{\Cov}{{\mathrm{Cov}}}
\newcommand{\cov}{{\mathrm{cov}}}
\newcommand{\GF}{{\mathrm {GF}}}
\newcommand{\mod}{{\mathrm mod\:}}
\newcommand{\rank}{{\mathrm{rank\:}}}
\newcommand{\spn}{{\mathrm{span\:}}}
\newcommand{\E}{{\mathrm E}}
\newcommand{\bmu}{\mbox{\boldmath $\mu$}}
\newcommand{\bxi}{\mbox{\boldmath $\xi$}}
\newcommand{\cL}{{\cal L}}
\newcommand{\bN}{{\bf N}}
\newcommand{\cB}{{\cal B}}
\newcommand{\cH}{{\cal H}}
\newcommand{\cU}{{\cal U}}
\newcommand{\baa}{\begin{eqnarray*}}
\newcommand{\eaa}{\end{eqnarray*}}

\newcommand{\ds}{{\:d\bf s}}
\newcommand{\du}{{\:d\bf u}}
\newcommand{\dy}{{\:d\bf y}}
\newcommand{\dH}{{\:d\bf H}}
\newcommand{\dHyus}{\dH\dy\du\ds}
\newcommand{\diag}{\textrm{diag}}

\newcommand{\bs}{{\bf s}}
\newcommand{\ba}{{\bf a}}
\newcommand{\bb}{{\bf b}}
\newcommand{\bq}{{\bf q}}
\newcommand{\qstar}{{q^{\star}}}
\newcommand{\Qstar}{{Q^{\star}}}
\newcommand{\xstar}{{x^{\star}}}
\newcommand{\ystar}{{y^{\star}}}
\newcommand{\bp}{{\bf p}}
\newcommand{\bX}{{\bf X}}
\newcommand{\obX}{{\bar {\bf X}}}
\newcommand{\obx}{{\bar {\bf x}}}
\newcommand{\obm}{{\bar {\bf m}}}
\newcommand{\obY}{{\bar {\bf Y}}}
\newcommand{\oby}{{\bar {\bf y}}}
\newcommand{\obZ}{{\bar {\bf Z}}}
\newcommand{\obU}{{\bar {\bf U}}}
\newcommand{\obW}{{\bar {\bf W}}}
\newcommand{\obu}{{\bar {\bf u}}}
\newcommand{\obw}{{\bar {\bf w}}}
\newcommand{\obN}{{\bar {\bf N}}}
\newcommand{\obM}{{\bar {\bf M}}}
\newcommand{\obB}{{\bar {\bf B}}}
\newcommand{\oX}{{\bar {X}}}
\newcommand{\oY}{{\bar {Y}}}
\newcommand{\ou}{{\bar {u}}}
\newcommand{\ow}{{\bar {w}}}
\newcommand{\oR}{{\bar {R}}}
\newcommand{\oM}{{\bar {M}}}
\newcommand{\oB}{{\bar {B}}}
\newcommand{\bU}{{\bf U}}
\newcommand{\bW}{{\bf W}}
\newcommand{\bY}{{\bf Y}}
\newcommand{\bV}{{\bf V}}
\newcommand{\bZ}{{\bf Z}}
\newcommand{\bT}{{\bf T}}
\newcommand{\bS}{{\bf S}}
\newcommand{\bM}{{\bf M}}
\newcommand{\bH}{{\bf H}}
\newcommand{\DFT}{{\mathrm{DFT}}}
\newcommand{\IDFT}{{\mathrm{IDFT}}}
\newcommand{\ldeg}{{\mathrm ldeg}}
\newcommand{\Real}{{\mathrm Re}}
\newcommand{\weight}{{\mathrm weight}}
\newcommand{\xor}{\oplus}
\newcommand{\bu}{{\bf u}}
\newcommand{\bv}{{\bf v}}
\newcommand{\bt}{{\bf t}}
\newcommand{\bd}{{\bf d}}
\newcommand{\bD}{{\bf D}}
\newcommand{\bw}{{\bf w}}
\newcommand{\bn}{{\bf n}}
\newcommand{\bx}{{\bf x}}
\newcommand{\by}{{\bf y}}
\newcommand{\bz}{{\bf z}}
\newcommand{\bone}{{\bf 1}}
\newcommand{\bzr}{{\bf 0}}
\newcommand{\cA}{{\cal A}}
\newcommand{\cP}{{\cal P}}
\newcommand{\cE}{{\cal E}}
\newcommand{\cF}{{\cal F}}
\newcommand{\cR}{{\cal R}}
\newcommand{\cS}{{\cal S}}
\newcommand{\cT}{{\cal T}}
\newcommand{\cX}{{\cal X}}
\newcommand{\cY}{{\cal Y}}
\newcommand{\oS}{\overline{S}}
\newcommand{\oP}{\overline{P}}
\newcommand{\hP}{\hat{P}}
\newcommand{\hR}{\hat{R}}
\newcommand{\tR}{\tilde{R}}
\newcommand{\hbF}{\hat{\bf F}}
\newcommand{\hF}{\hat{F}}
\newcommand{\hbU}{{\bf{\hat{U}}}}
\newcommand{\tbU}{{\bf{\tilde{U}}}}
\newcommand{\hbS}{{\bf{\hat{S}}}}
\newcommand{\hw}{\hat{w}}
\newcommand{\cC}{{\mathcal{C}}}
\newcommand{\cN}{{\mathcal{N}}}
\newcommand{\cG}{\mathcal{G}}
\newcommand{\hcC}{\hat{\mathcal{C}}}
\newcommand{\cD}{\mathcal{D}}
\newcommand{\hcD}{\hat{\mathcal{D}}}
\newcommand{\hD}{\hat{D}}
\newcommand{\tcD}{\tilde{\mathcal{D}}}
\newcommand{\tcC}{\tilde{\cal C}}
\newcommand{\tC}{\tilde{C}}
\newcommand{\hA}{\hat{A}}
\newcommand{\hB}{\hat{B}}
\newcommand{\hC}{\hat{C}}
\newcommand{\hc}{\hat{c}}
\newcommand{\btc}{\tilde{\bf c}}
\newcommand{\hbr}{\hat{\bf r}}
\newcommand{\hbw}{\hat{\bf w}}
\newcommand{\hbc}{\hat{\bf c}}
\newcommand{\hby}{\hat{\bf y}}
\newcommand{\hX}{\hat{X}}
\newcommand{\hY}{\hat{Y}}
\newcommand{\hy}{\hat{y}}
\newcommand{\hbx}{\hat{\bf x}}
\newcommand{\hbm}{\hat{\bf m}}
\newcommand{\hbX}{\hat{\bf X}}
\newcommand{\hbY}{\hat{\bf Y}}
\newcommand{\beginproof}{\noindent \textbf{Proof: }  }
\newcommand{\finproof}{\noindent $\Box$\\}
\newcommand{\ve}{\varepsilon}
\newcommand{\emptyline}{$\:\\ $}
\def\refeq#1{\: {\stackrel{ (#1)}{=}} \: }
\def\defined{\: {\stackrel{\scriptscriptstyle \Delta}{=}} \: }
\def\psdir#1{#1}
\def\MSE{{\rm MSE}}
\def\argmax{\mathop{\rm argmax}}
\def\CB{\mathop{\rm BS}}
\def\defined{\: {\stackrel{\scriptscriptstyle \Delta}{=}} \: }
\def\leqa{\buildrel \rm {\scriptscriptstyle (1)} \over \leq}
\def\leqb{\buildrel \rm {\scriptscriptstyle (2)} \over \leq}
\def\leqc{\buildrel \rm {\scriptscriptstyle (3)} \over \leq}
\def\eqa{\buildrel \rm {\scriptscriptstyle (1)} \over =}
\def\eqb{\buildrel \rm {\scriptscriptstyle (2)} \over =}
\def\eqc{\buildrel \rm {\scriptscriptstyle (3)} \over =}
\newfont{\boldlarge}{msbm10 scaled 1100}
\newcommand{\RR}{\mbox{\boldlarge R}}
\newcommand{\tr}{\mathrm{tr}}

\newcommand{\eqref}[1]{(\ref{#1})}
\newcommand{\labeleq}[1]{\label{eq:#1}}
\newcommand{\refsec}[1]{(\ref{section:#1})}
\newcommand{\labelsec}[1]{\label{section:#1}}

\newcommand{\comment}[1]{}
\newcommand{\topc}[1]{$\stackrel{\circ}{\rm #1}$}
\newcommand{\bbb}[1]{$<$ \textbf{To be completed: #1} $>$}
\newcommand{\ShowDavid}[1]{$<$ \textbf{Show David: #1} $>$}
\newcommand{\ShowAmir}[1]{$<$ \textbf{Show Amir: #1} $>$}
\newcommand{\ShowDavidNew}[1]{$<$ \textbf{Show David New: #1} $>$}
\newcommand{\ShowAmirNew}[1]{$<$ \textbf{Show Amir New: #1} $>$}
\newcommand{\ShowDavidNewNew}[1]{$<$ \textbf{Show David 2: #1} $>$}
\newcommand{\Kavcic}{Kav\u{c}i\'{c}}

\def\etal{$et \,\,al.\,\,$}

\renewcommand{\thesection}{\Roman{section}}

\newlength{\tmpbigbar}

\newcommand{\bigbar}[1]{
 \setlength{\tmpbigbar}{\unitlength}
 \settowidth{\unitlength}{\mbox{$#1$}}
  \stackrel{\barpic}{#1}
 \setlength{\unitlength}{\tmpbigbar}
}
\newcommand{\barpic}{\begin{picture}(1,0.01)(0,0)
\put(0.15,0){\line(12,0){0.7}}
\end{picture}
}

\title{On the Fading Paper Achievable Region of the Fading MIMO Broadcast Channel
\thanks{Manuscript submitted to IEEE Transactions on Information Theory: June 2006, revised: July 2007, to be published.  This research was supported by the Israel Science Foundation, grant no. 927/05.
The material in this paper was presented at the 44th Annual
Allerton Conference on Communications, Control and Computing,
Monticello, IL, September 2006. } }

\author{Amir~Bennatan,~\IEEEmembership{Member,~IEEE}\thanks{A.~Bennatan was with the School of
Electrical Engineering, Tel-Aviv University, Tel-Aviv 69978,
Israel. He is now with the Program in Applied and Computational
Mathematics (PACM) at Princeton University, Princeton, NJ 08544
USA (e-mail: abn@math.princeton.edu).}
and~David~Burshtein,~\IEEEmembership {Senior
Member,~IEEE}\thanks{D.~Burshtein is with the School of Electrical
Engineering, Tel-Aviv University, Tel-Aviv 69978, Israel (e-mail:
burstyn@eng.tau.ac.il). }}

\markboth{To be published, IEEE Transactions on Information
Theory} {Bennatan and Burshtein: On the Fading Paper Achievable
Region of the Fading MIMO Broadcast Channel}

\maketitle \setcounter{page}{1}

\begin{abstract}
We consider transmission over the ergodic fading multi-antenna
broadcast (MIMO-BC) channel with partial channel state information
at the transmitter and full information at the receiver. Over the
equivalent {\it non}-fading channel, capacity has recently been
shown to be achievable using transmission schemes that were
designed for the ``dirty paper'' channel.  We focus on a similar
``fading paper'' model.  The evaluation of the fading paper
capacity is difficult to obtain.  We confine ourselves to the {\it
linear-assignment} capacity, which we define, and use convex
analysis methods to prove that its maximizing distribution is
Gaussian. We compare our fading-paper transmission to an
application of dirty paper coding that ignores the partial state
information and assumes the channel is fixed at the average fade.
We show that a gain is easily achieved by appropriately exploiting
the information.  We also consider a cooperative upper bound on
the sum-rate capacity as suggested by Sato.  We present a numeric
example that indicates that our scheme is capable of realizing
much of this upper bound.
\end{abstract}


\section{Introduction} \label{sec:Introduction}
The multiple-antenna Gaussian broadcast channel has recently been
the subject of intense research.  This surge of interest was
spurred by the seminal work of Caire and
Shamai~\cite{Caire_Shamai_MIMO}, who suggested an achievable
region for this channel based on dirty-paper coding. Recently,
this region was shown by Weingarten~\etal~\cite{HananYossiShlomo}
to exhaust the capacity region of the channel.

However, the channel model examined in~\cite{Caire_Shamai_MIMO}
assumes that the fading coefficients of the MIMO channel are fixed
and known to both the transmitter and the receiver.  In several
realistic settings, the coefficients fluctuate over time.  They
are estimated at the receiver and are fed back to the transmitter.
At best, we can assume that the transmitter has a rough, outdated
estimate of the coefficients.

Telatar~\etal~\cite{Telatar}, in his work on the single-user MIMO
channel, focused on a setting where the transmitter has zero
knowledge of the fading coefficients.  In a broadcast setting,
this problem is typically uninteresting because its solution is
often trivial. In Appendix~\ref{apdx:TDMA}, we will see such a
setting where time-sharing (TDMA) is the best that can be
achieved. However, in a realistic setting, the transmitter has
{\it some} knowledge of the channel to each of the users. This
knowledge can be modelled as channel distribution
information\footnote{A different model was proposed by
Jindal~\cite{Jindal} and Caire~\cite{Caire_Talk}, who incorporated
the feedback from the receiver into the channel model. }.

We assume an ergodic channel, in the sense that a new channel
realization is obtained at each time instance.  However, the
channel distribution, which is known to the transmitter, remains
fixed for the duration of the transmission.

The analysis of ergodic broadcast channels was initiated by
Cover~\cite{Cover_Broadcast}.  The capacity of such channels is
known only in special cases, where the signals to the users can be
ordered according to their ``strength''.  A large class of such
channels, known as ``more capable'' channels, was considered by El
Gamal~\cite{More capable}, who also evaluated the capacity in this
case.  This class contains ``degraded'' and ''less noisy''
channels as special cases~\cite{More capable}.

Tuninetti and Shamai~\cite{Tuninetti} considered the fading {\it
scalar} broadcast channel, which is a special case of the fading
MIMO-BC channel obtained by setting the number of antennas at the
transmitter and receivers to one.  They showed that this channel
is not ``more capable'' in general.  They nonetheless evaluated
the ``more capable'' region as defined by~\cite{More capable}.
This region is still achievable despite the channel being not
``more capable'', although it is only an inner bound and does not
exhaust the entire capacity region.

Jafar~\etal~\cite{Goldsmith_Isotropic} considered the fading
MISO-BC, characterized by receivers that have only one antenna
each.  They considered the case when the distribution of the
fading coefficients is isotropic.  In this case, they proved that
the capacity region collapses to that of the above fading scalar
channel. Lapidoth~\cite{Lapidoth_MIMO} examined a similar two-user
fading MISO-BC channel, and demonstrated that at the limit of high
SNR, a significant loss is incurred as a result of the
unavailability of precise channel state information at the
transmitter.  Sharif and Hassibi~\cite{Hassibi} proposed a
beamforming transmission approach for the case when the knowledge
available to the transmitter is the collection of SINR values
available to each of the receivers.

The fading MIMO-BC channel, being not ``more capable'' in general,
is difficult to analyze.  In this paper we focus on an achievable
region which is modelled on the dirty paper region of Caire and
Shamai~\cite{Caire_Shamai_MIMO}. Our development uses a {\it
fading-paper} approach which is a generalization of the
dirty-paper approach of~\cite{Caire_Shamai_MIMO}.
A fading paper solution was previously considered for a wideband fading channel
in~\cite{fading_paper_causal}, although they assumed an interference which is
known only causally, unlike the dirty paper problem of Costa.
The proof of~\cite{HananYossiShlomo} does not apply to the fading
MIMO-BC capacity region, so that the fading
paper approach is not guaranteed to be optimal. Furthermore, the capacity of the
fading-paper channel is in general not known.  We focus on its
{\it linear-assignment} capacity, which we define.  We use
convex-optimization methods to prove that a Gaussian distribution
achieves this capacity.

We compare the rate region achieved by this approach to the region
that is achievable by a dirty-paper scheme that ignores the
available channel state information and assumes that the channel
is fixed at its average. We show that a substantial benefit is
easily achieved by appropriately exploiting the available
information.

This paper is organized as follows.  We begin with some background
in Sec.~\ref{sec:background}.  We define our notation and the
channel model, discuss the dirty-paper channel and its application
to transmission over the non-fading MIMO-BC channel.  In
Sec.~\ref{sec:Fading_Paper} we discuss the fading-paper
generalization of the dirty-paper channel, define the
linear-assignment capacity and discuss its maximizing
distribution.  In Sec.~\ref{sec:Fading_Paper_Achievable} we define
a region that is achievable using linear-assignment fading-paper
transmission methods. We also compare this region to that of
dirty-paper based transmission that assumes the channel is fixed
at its average. In Sec.~\ref{sec:Conclusion} we present ideas for
further research and conclude the paper.
\section{Background} \label{sec:background}
\subsection{Notation} \label{sec:notation}
$\E_{H}$ denotes the expectation over the random variable $H$.
Matrices are denoted by upper-case letters, with bold indicating
realizations of random variables (e.g. $\bH$ is the realization of
$H$).  Vector values are denoted in boldface and scalar values are
denoted in normal typeface.  With both, lower-case letters denote
the realizations of random variables ($\by$ is a realization of
$\bY$ and $y$ is a realization of $Y$).

The inner product of two equal-dimension matrices $A, B \in
\RR^{M\times N}$ is defined by,
\begin{eqnarray*}
<A,B> \defined \sum_{m=1}^M\sum_{n=1}^N A_{m,n}B_{m,n} =
\tr[A\cdot B^T]
\end{eqnarray*}
$\RR_{+}$ denotes the non-negative real numbers and $\RR_{++}$ the
positive real numbers.
\subsection{System Model} \label{sec:channel model} We consider a
broadcast channel with $L$ users.  The transmitter has $M$
transmit antennas and user $l$ has $N_l$ antennas.   For
simplicity we assume that all signals are real-valued.

The channel output $\bY^{(l)}_t$ observed by receiver $u$ at a
discrete time instance $t$ is given by,
\begin{eqnarray*}
\bY^{(l)}_t = H^{(l)}_t\cdot \bX_t + \bZ^{(l)}_t
\end{eqnarray*}
 $\bY^{(l)}_t$ is a $N_l \times 1$ column vector.  $\bH^{(l)}_t$ is
 a random $N_l \times M$ matrix denoting the channel transition matrix.
 We assume that instances of $H^{(l)}_t$ are independent over
 time (for different values of $t$) and between users (i.e., for
 different values of $l$).  As noted in Sec.~\ref{sec:Introduction},
 we assume that this matrix is known to the receiver, and in our
 subsequent analysis, we consider it as part of the channel
 output.  $\bX_t$ is an $M \times 1$ column vector denoting the transmitted
 signal. $\bZ^{(l)}_t$ denotes Gaussian noise, distributed as a $N_l$-dimensional
 zero-mean Gaussian random variable with identity covariance matrix $\bI$~\footnote{If
 the noise's covariance matrix is not $\bI$, we can multiply $\bY^{(l)}_t$ by the inverse
 of the square root of of the matrix and obtain an equivalent channel that does
 agree with this model.}.

In the sequel, for simplicity, we will drop the time index $t$. We
assume that the transmitter is subject to an average power
constraint $P$.  That is, we require,
\begin{eqnarray*}
\E\: \tr(\bX\bX^T) \leq P
\end{eqnarray*}
The only assumption we make on the distribution of $H^{(l)}$ is
that is has finite energy, i.e. $\E <H^{(l)}, H^{(l)}>$ is finite.
\subsection{Dirty Paper Channels}\label{sec:Dirty_Paper}
The dirty-paper channel was first considered by
Costa~\cite{costa:83}. It is defined by
\begin{eqnarray}  \label{eq:DirtyPaper}
Y = X + S + Z
\end{eqnarray}
The channel input $X$ is subject to a power constraint $P$, i.e.
The noise $Z$ is distributed as a zero-mean Gaussian variable with
variance $\sigma_Z^2 > 0$. $S$ is interference, known to the
transmitter but not to the receiver.

Costa obtained the remarkable result that the interference,
despite being known only to the encoder, incurs no loss of
capacity in comparison with the standard interference-free
channel.  Costa assumed that $S$ is Gaussian i.i.d distributed.
This result was extended in \cite{cohen:2002} and
\cite{esz:Lattice_Known_Interference} to arbitrarily distributed
interference. Costa's result was further extended to the Gaussian
MIMO channel by Yu~\etal~\cite{Yu_Colored_Paper}. With this
channel model, vector $\bY$, $\bS$, $\bX$ and $\bZ$ replace the
above scalar equivalents, $\bZ$ being a zero-mean Gaussian random
vector with nonsingular covariance matrix
$\Sigma_Z$~\footnote{Note that unlike the fading MIMO-BC model of
Sec.~\ref{sec:Fading_Model}, we find it more convenient to allow
$\Sigma_Z \neq \bI$ in this context of the vector dirty-paper
channel. }.

In Sec.~\ref{sec:Dirty_Paper_Achievable} we will consider
dirty-paper in the context of transmission over nonfading MIMO-BC
channels. In that context, it will be useful to consider the
following variation of~\eqref{eq:DirtyPaper} (using vector
substitutes for $Y$, $S$, $X$ and $Z$),
\begin{eqnarray}  \label{eq:MIMO_DirtyPaper}
\bY = \bH(\bX + \bS) +\bZ
\end{eqnarray}
where $\bS$ and $\bX$ are $M$ dimensional, $\bY$ and $\bZ$ are $N$ dimensional, and $\bH$ is an $N \times M$ fixed channel matrix\footnote{The matrix $\bH$ is denoted in bold since in the next section it will be a realization of a random variable.}. We assume
this formulation of the dirty-paper problem throughout the rest of
this paper.  Once again, the capacity coincides with that of the
corresponding no-interference channel, whose output $\hbY$ is
given by,
\begin{eqnarray}\label{eq:No_Interference}
\hbY = \bH\bX + \bZ
\end{eqnarray}
The dirty-paper channel is an instance of the more general class
of {\it side-information} channels, first considered by
Shannon~\cite{shannon:58}.  Such channels are characterized by an
input $X$, output $Y$ and state-dependent transition probabilities
$\Pr[y|x,s]$ where the channel state $S$ is i.i.d., known to the
transmitter and unknown to the receiver.  In the context
of~\eqref{eq:DirtyPaper}, the interference $S$ constitutes the
channel state.

Shannon~\cite{shannon:58} considered the case of the state
sequence being known only causally.  Kusnetsov and
Tsybakov~\cite{kustsy:74} were the first to consider the case of
state sequence known non-causally, and Gel'fand and
Pinsker~\cite{gelpinsk:80} obtained the capacity formula for this
case.  The capacity of this channel is given by
\begin{equation} \label{eq:GelPinsk}
C = \sup_{\Pr[u\given s],\rmf(\cdot)} \{ I(U;Y) - I(U;S) \}
\end{equation}
where $U$ is an auxiliary random variable with conditional
distribution $\Pr[u\given s]$ and $\rmf(\cdot)$ is a deterministic
function, such that the transmitted signal $X$ is given by $X =
\rmf(S,U)$.

In~\cite{Yu_Colored_Paper}, the capacity of the dirty-paper
channel was obtained from \eqref{eq:GelPinsk} using an auxiliary
random variable $\bU$ given by  $\bU = \bF\cdot \bS + \bX$, where
$\bF$ is a fixed matrix\footnote{We denote the matrix $\bF$ in bold throughout the paper in order to distinguish it from the functional $F(q,Q)$.} and $\bX$ is a zero-mean
Gaussian-distributed random-variable, independent of $\bS$.  The
use of $\bX$ has a dual role.  First, it is a component in the
definition of the transition probabilities $\Pr[\bu \given \bs]$.
Second, given $\bU$ and $\bS$, the transmitted signal satisfies
$\bof(\bU, \bS) \defined \bU - \bF\cdot \bS = \bX$.  The
covariance matrix $\Sigma_X$ of $\bX$ is determined as in the
no-interfence channel (see e.g.~\cite{Cover_Book}).  An expression
for $\bF$ was developed by Yu and Cioffi~\cite{Yu_MIMO_BC}.  In
this paper, we use the following, equivalent expression:
\begin{eqnarray}\label{eq:F_DPC}
\bF = \Sigma_{X}\bH^T(\bH\Sigma_{X}\bH^T + \Sigma_Z)^{-1}\bH
\end{eqnarray}
A proof that this choice of $\bF$ indeed achieves the
no-inteference capacity is provided in
Appendix~\ref{apdx:Proof_F}.  This proof is different from the
proof of~\cite{Yu_MIMO_BC}, and is provided primarily for
completeness.

Costa~\cite{costa:83} and Yu~\cite{Yu_Colored_Paper} obtained
their results using random codes and maximum-likelihood decoding.
Zamir~\etal~\cite{esz:it02} and
Bennatan~\etal~\cite{SuperpositionCoding} have presented practical
methods for transmitting at rates that approach the above computed
capacities.  Their approaches were developed for the scalar
dirty-paper channel, but can easily be adapted to the MIMO setting
\cite{SuperpositionCoding}[Sec.~VII].
\subsection{The Dirty-Paper Achievable Region}\label{sec:Dirty_Paper_Achievable}
In their construction for the non-fading MIMO broadcast channel,
Caire and Shamai~\cite{Caire_Shamai_MIMO} used dirty-paper coding
to transmit in the following way.  The transmitted signal $\bX$ is
constructed as the vector sum of $L$ signals $\bX_1,...,\bX_L$,
where $\bX_l$ contains the transmitted signal to user $l$. Each
user is also allotted a virtual power constraint $P_l$ such
that $\sum_{l=1}^L P_l = P$. Using dirty-paper coding, the
transmitter can generate the signal $\bX_l$ such that the
interference generated by $\bX_1,...,\bX_{l-1}$ is effectively
pre-subtracted. More precisely, encoding proceeds in the following
way,
\begin{enumerate}
\item The transmitter begins by selecting a codeword $\bc_1$ for user 1.
\item It then proceeds to determine the signal for user 2.  It constructs the
signal $\bX_2$ for user 2 using a dirty-paper transmission scheme,
making use of its full non-causal knowledge of $\bc_1$ and
treating it as known interference (in lieu of $S$ in
\eqref{eq:DirtyPaper}).
\item The signals $\bX_3,...,\bX_L$ are constructed in a similar
manner.  When constructing the signal to user $l$, the signal
$\bS^{(l)}\defined \bX_1+\bX_2+...+\bX_{l-1}$ is treated as
non-causally known interference.
\end{enumerate}
The operation of the receivers mirrors the above transmission
scheme.  Receiver $l$ applies dirty-paper decoding, effectively
cancelling the interference generated by
$\bX_1+\bX_2+...+\bX_{l-1}$ but treating $\bX_{l+1}+...+\bX_{L}$
as part of the unknown noise (alongside $\bZ$).

The above transmission strategy defines an achievable rate region
for the Gaussian MIMO broadcast channel.  This region is a
function of the virtual power constraints $P_l$ imposed on the
users.  Furthermore, it is a function of the covariance matrices
$\Sigma_{X}^{(l)}$ by which the various codebooks for the signals
$\bX_l$ are randomly generated. It is also a function of the
ordering of the users. The convex-hull of the union of all regions
obtained in this way constitutes the dirty-paper achievable region
$\cC_{DPC}(P)$. In~\cite{HananYossiShlomo}, this region was shown
to exhaust the MIMO broadcast capacity region.

However, the application of dirty-paper transmission methods in
the above algorithm is heavily reliant on the availability of
precise knowledge of the fixed channel matrices
$\{\bH^{(l)}\}_{l=1}^{L}$ at the transmitter.  Without these, the
pre-subtraction of the signals $\{\bX_i\}_{i < l}$, when
constructing $\bX_l$, is not possible.

\section{The Fading-Paper Problem}\label{sec:Fading_Paper}

\subsection{Channel Model}\label{sec:Fading_Model} The fading-paper
channel is an adaptation of the dirty-paper model (as expressed
in~\eqref{eq:MIMO_DirtyPaper}) of Sec.~\ref{sec:Dirty_Paper},
designed to account for the absence of channel state information
at the receiver.  The channel is defined by,
\begin{eqnarray}  \label{eq:FadingDirtyPaper}
\bY = H(\bX + \bS) + \bZ
\end{eqnarray}
Unlike the case in~\eqref{eq:MIMO_DirtyPaper}, the channel matrix
is random and is know to the receiver but not to the transmitter.
The pair $(\bY,H)$ constitutes the channel output, where $\bY$ is
the channel observation and $H$ is the channel matrix.

The channel transition probabilities are also a function of the
distribution of the interference $\bS$ and of the channel matrix
$H$. In this paper, we assume $\bS$ to be a zero-mean Gaussian
distributed random variable with covariance $\Sigma_S$.  As noted
in Sec.~\ref{sec:channel model}, we make no assumptions on the
distribution of $H$, beyond it having finite energy. Following the
discussion of side-information channels in
Sec.~\ref{sec:Dirty_Paper}, the capacity of the fading-paper
channel is given by,
\begin{equation} \label{eq:CapFadingDirtyPaper}
C = \sup_{\Pr[\bu\given \bs],\bof(\cdot)} \{ I(\bU;\bY,H) -
I(\bU;\bS) \}
\end{equation}
where $\bU$ is an auxiliary random variable whose joint
distribution with $\bS$ can be obtained via $\Pr[\bu\given \bs]$.
$\bof(\cdot)$ is a vector-valued deterministic function, such that
the transmitted signal $\bX$ is given by $\bX = \bof(\bU,\bS)$.

Note that for any particular choice of $\Pr[\bu\given \bs]$ and
$\bof(\cdot)$, the contents of the braces are an achievable
transmission rate over the channel,
\begin{eqnarray}\label{eq:FadingAchievable}
R_{\textrm{achievable}} = I(\bU;\bY,H) - I(\bU;\bS)
\end{eqnarray}
\subsection{The Linear-Assignment Capacity}\label{sec:Fading_Lin_Strat}
In this paper, we focus on a subset of achievable rates for the
fading-paper channels, modelled on the dirty-paper
capacity-achieving assignment for $\bU$ and $\bof(\cdot)$.  That
is, we focus on an auxiliary random variable $\bU$ given by
\begin{equation}
\label{eq:F}
\bU = \bF\cdot \bS + \bX
\end{equation}
where $\bF$ is some arbitrary real-valued $M\times M$ matrix, and
$\bX$ is an arbitrary zero-mean random-variable, which may depend
on $\bS$. We define $\bof(\bu, \bs) = \bu - \bF\bs$. We refer to
such an assignment as a {\it linear assignment}. We call the
maximum in~\eqref{eq:CapFadingDirtyPaper}, when restricted to such
assignments, the {\it linear assignment capacity}.

Linear assignments may equivalently be defined as follows. A
linear assignment is characterized by an arbitrary zero-mean
$M$-dimensional random variable $\bU$ (recall that $M$ is the
dimension of $\bX$ and $\bS$), which may be dependent on $\bS$,
and an arbitrary real-valued $M\times M$ matrix $\bF$.  In the
context of~\eqref{eq:GelPinsk}, $\bU$ corresponds to the auxiliary
variable $\bU$ and $\bof(\cdot,\cdot)$ is defined by
$\bof(\bu,\bs) = \bu - \bF\bs$. A set $\bU, \bF$ and
$\bof(\cdot,\cdot)$ given by the first definition
straightforwardly satisfies the conditions of the second
definition.  To see that the reverse holds, observe that we have
allowed $\bX$  to be completely arbitrary.  In particular, we have
in no way required $\bX$ to be Gaussian or independent of $\bS$.
Thus, given a pair $\bU$ and $\bF$ corresponding to the second
definition, we may define $\bX = \bU - \bF\cdot\bS$ and the
resulting set $\bU, \bX, \bF$ and $\bof(\cdot,\cdot)$ coincides
with the first definition.

The optimality of linear assignments for the dirty-paper problem
of Sec.~\ref{sec:Dirty_Paper} is obtained from the fact that their
maximum achievable rate coincides with the capacity of the
corresponding no-interference channel.  This is clearly the best
we can hope for, and thus such assignments achieve capacity. With
fading-paper, the achievable rate with linear assignments is in
general strictly {\it below} the no-interference upper-bound.
Thus, it is not known whether it is optimal.

In our above definition of linear assignments, we left the
distribution of $\bX$ undefined.  Specifically (as noted above),
we did not insist on $\bX$ to be Gaussian, and did not insist on
it being independent of $\bS$, as we did in
Sec.~\ref{sec:Dirty_Paper} when we discussed the
capacity-achieving assignment for the dirty-paper channel.
However, the following theorem establishes the optimality of a
Gaussian-distributed $\bX$.  In
Sec.~\ref{sec:Fading_Paper_Achievable} we will show that we may
also assume $\bX$ to be independent of $\bS$.

In the following theorem, we assume the following regularity conditions:
\begin{enumerate}
\item We assume that the
expectations~\eqref{eq:27},~\eqref{eq:28},~\eqref{eq:29}
and~\eqref{eq:30} (defined below), exist and are finite.  Note
that this condition is satisfied, for example, when the
distribution of $H$ is discrete and takes a finite set of values.
\item We assume that the covariance matrix of the vector
$(\bS,\bU)$,
\begin{equation}
\label{eq:sigma_su}
\left(
  \begin{array}{cc}
    \Sigma_S       & \Sigma_{S,U} \\
    \Sigma_{S,U}^T & \Sigma_U \\
  \end{array}
\right)
\end{equation}
is nonsingular (i.e., it is a positive definite matrix). Note that
this also implies that $\Sigma_S$ is nonsingular, being a
principal submatrix of $\Cov(\bU,\bS)$. Since,
$$
\left(
  \begin{array}{c}
    \bS \\
    \bU \\
  \end{array}
\right) =
\left(
  \begin{array}{cc}
    I & 0 \\
    F & I \\
  \end{array}
\right)
\left(
  \begin{array}{c}
    \bS \\
    \bX \\
  \end{array}
\right)
$$
and since the matrix on the right hand side of the last equation
is nonsingular, a sufficient condition that~(\ref{eq:sigma_su}) is
nonsingular is that $\det(\Sigma_S)>0$, $\det(\Sigma_X)>0$ and
$\det(\Sigma_X - \Sigma_{S,X}^T \Sigma_S^{-1} \Sigma_{S,X} ) > 0$
($\Sigma_X$ and $\Sigma_{S,X}$ are the covariance of $\bX$ and the
cross-covariance of $\bS$ and $\bX$,
respectively).
\item
We assume an arbitrary density $q(\bu\given\bs)$ with respect to the Lebesgue measure.
\end{enumerate}
\begin{definition}\label{def:setting}
Given a linear assignment, the collection of matrices $\Sigma_S,
\Sigma_{S,X}, \Sigma_X$ and $\bF$ is called its {\it setting}.
\end{definition}
\begin{theorem}\label{theorem:Gaussian_Maximizes}
Assume the above-mentioned regularity conditions. For any fixed setting, the linear-assignment capacity (as defined
above) is achieved by a choice of $\bX$ that is jointly Gaussian with $\bS$.
\end{theorem}
\begin{proof}
We begin with a brief outline of the proof. We
consider~\eqref{eq:FadingAchievable} as a function of the density
$q(\bu\given\bs)$ and of $Q(\bu\given\by,\bH)$, defined below.  We
then seek to show that $q_G(\cdot)$ and $Q_G(\cdot)$,
corresponding to a joint-Gaussian choice of $\bX$ and $\bS$,
maximize~\eqref{eq:FadingAchievable}.  To do so, we pose the
problem as a concave constrained maximization problem, and show
that $q_G$ and $Q_G$ admit Lagrange multipliers.

We now rewrite~\eqref{eq:FadingAchievable} as $F(q,Q)$, given
by\footnote{This definition is an adaptation of a similar
definition by Heegard and El Gamal~\cite{Heegard_Side_Inf}},
\begin{eqnarray}
\nonumber
F(q,Q) &\defined& \int_{\bs\in\RR^M}\int_{\bu\in\RR^M}\int_{\by\in
\RR^N}\int_{\bH\in\cR_H}f_{\bS}(\bs)f_{Y,H\given
S,X}(\by,\bH\given \bs, \bx = \bof(\bu,
\bs))q(\bu\given\bs) \cdot \\
& & \cdot \log\frac{Q(\bu\given\by,\bH)}{q(\bu\given\bs)}\dH
\dy \du \ds
\label{eq:F_q_Q_defined}
\end{eqnarray}
Recall that $M$ and $N$ are the dimensions of $\bS$ and $\bY$, respectively. We also denote by
$\cR_H$ the support region of the random variable $\bH$.
$Q(\bu\given\by,\bH)$ is the conditional distribution of the
above-defined $\bU$ given the channel output $\bY$ and the signal
fade $\bH$. $f_{S}(\bs)$ is the density of $\bS$ and
$f_{Y,H\given S,X}(\by,\bH\given \bs, \bx)$ is the conditional
density of $\bY$ and $\bH$ given the transmitted $\bx$ and
interference $\bs$.

Since we make no assumptions on the distribution of $\bH$, the
existence of this density is not guaranteed.  However, the
generalization to the case when the density does not exist is
straightforward. In the sequel, we drop the subscripts and denote
the densities by $f(\bs)$ and $f(\by,\bH\given \bs, \bx)$.  Note
that $f(\bs)$ should not be confused with the previously defined
$\bof(\bu, \bs)$.

We defined $Q(\bu\given\by,\bH)$ in~\eqref{eq:F_q_Q_defined} to be
the conditional density of the above-defined $\bU$ given the
channel output $\bY$ and the signal fade $\bH$. Actually, in the
sequel we find it convenient to relax this requirement and
consider $F(q,Q)$ for arbitrary probability densities
$Q(\bu\given\by,\bH)$. However, the pair $q$ and $Q$ that {\it
maximizes} $F(q,Q)$ {\it will} satisfy the requirement.  In this
we follow the example of~\cite{Heegard_Side_Inf}.

For given $\Sigma_S$, $\Sigma_X$ and $\Sigma_{S,X}$, let
$q_G(\bu\given\bs)$ and $Q_G(\bu\given\by,\bH)$ denote the
conditional densities corresponding to the choice of $\bX$ that is
jointly-Gaussian with $\bS$. Our objective is to show that $q_G$
and $Q_G$ maximize $F(q,Q)$.

$F(q,Q)$ as defined in~\eqref{eq:F_q_Q_defined} is jointly-concave
in its arguments.  Thus we may wish to apply methods from the
theory of convex optimization to maximize it.  Formally, we seek
to solve the following constrained problem
\begin{eqnarray}
\max_{q,Q} F(q,Q)\: \textrm{subject to}\label{eq:P}\\
\int_{\bs\in\RR^M}\int_{\bu\in\RR^M}f(\bs)q(\bu\given\bs)\left[\bof(\bu,
\bs)\cdot\bof(\bu, \bs)^T\right]\du\ds &=& \Sigma_X  \label{eq:E_q}\\
\int_{\bs\in\RR^M}\int_{\bu\in\RR^M}f(\bs)q(\bu\given\bs)\left[\bs\cdot\bof(\bu,
\bs)^T\right]\du\ds  &=& \Sigma_{S,X}  \label{eq:C_q}\\
\int_{\bu\in\RR^M} q(\bu\given\bs)\du &=& 1\quad\quad \forall\bs \in \RR^M \label{eq:q_valid}\\
\int_{\bu\in\RR^M} Q(\bu\given\by, \bH)\du &=& 1\quad\quad
\forall\by \in \RR^N,\forall \bH \in \cR_H \label{eq:Q_valid}
\end{eqnarray}
Recall that Theorem~\ref{theorem:Gaussian_Maximizes} assumes a
fixed setting. Thus, the matrices $\Sigma_S, \Sigma_{S,X},
\Sigma_X$ and $\bF$ are assumed to be given and fixed.  The
maximization is performed over the set of distributions
corresponding to these matrices, and our objective is to show that
a Gaussian distribution is optimal. Optimization of the matrices
themselves is beyond the scope of this proof (such optimization
will be discussed in Sec.~\ref{sec:Comparison Dirty Paper}).

\eqref{eq:E_q} and~\eqref{eq:C_q} are derived from the conditions
$\Sigma_X$ and $\Sigma_{S,X}$ on the transmitted signal $\bX$.
 That is, recalling
that $\bX = \bof(\bU,\bS)$, they are equivalent to
\begin{eqnarray*}
\E \left[\bX\cdot \bX^T\right] = \Sigma_X, \quad \E \left[\bS\cdot
\bX^T\right] = \Sigma_{S,X}
\end{eqnarray*}
To further simplify our analysis, we allow the arguments  $q$ and
$Q$ of $F(q,Q)$ to be arbitrary nonnegative measurable functions.
Constraints~\eqref{eq:q_valid} and~\eqref{eq:Q_valid}, compensate
for this and ensure that the final result is a valid conditional
distribution. Functions $q$ and $Q$ that satisfy constraints
\eqref{eq:E_q},~\eqref{eq:C_q},~\eqref{eq:q_valid}
and~\eqref{eq:Q_valid} are called {\it feasible}.

A straightforward approach to our optimization problem would
appear to be to apply the Karush-Kuhn-Tucker (KKT) conditions to
find the global maximum.  In reality, this is slightly more
involved because equations~\eqref{eq:q_valid}
and~\eqref{eq:Q_valid} involve an infinite number of constraints.
Furthermore, the arguments of $F(q,Q)$ are functions rather than
vectors. In~\cite{KKT_Necessity}, the necessity of the KKT
conditions was proven under certain conditions.  In this paper, we
only require their sufficiency for convex functionals, which is
easier to prove.  Our proof is tailored to the setting of our
particular problem. We begin by defining Lagrange multipliers.
\begin{definition}\label{def:Lagrangians}
Let $q$, $Q$ be two positive-valued\footnote{The condition that
$q$ and $Q$ be positive-valued is required for the expressions
that follow, which involve division by $Q(\bu\given\by,\bH)$ and
$q(\bu\given\bs)$, to be valid. } feasible functions. Lagrange
multipliers for $q$ and $Q$ are matrices $\Gamma,\Upsilon \in \RR^{M\times M}$,
and real-valued functions $\alpha(\bs):\RR^M \rightarrow \RR$ and
$\beta(\by,\bH):\RR^N\times\cR_H \rightarrow \RR$ such that,
\begin{eqnarray}
& &\int_{\by\in \RR^N}\int_{\bH\in\cR_H}f(\bs)f(\by,\bH\given \bs,
\bx = \bof(\bu,
\bs))\left[\log\frac{Q(\bu\given\by,\bH)}{q(\bu\given\bs)} -
1\right]\dH \dy +\nonumber\\& &\quad\quad\quad\quad
f(\bs)<\Upsilon,\bof(\bu, \bs)\cdot\bof(\bu,
\bs)^T>+f(\bs)<\Gamma,\bs\cdot\bof(\bu, \bs)^T>+\alpha(\bs) = 0
\nonumber\\& &
\quad\quad\quad\quad\quad\quad\quad\quad\quad\quad\quad\quad\quad\quad\quad\quad\quad\quad\quad\quad\quad\quad\quad\quad\quad\quad\quad\quad\quad
\forall\bs\in\RR^M, \forall\bu\in\RR^M\label{eq:Lagrange_q}\\
& &\int_{\bs\in \RR^M}f(\bs)f(\by,\bH\given \bs, \bx = \bof(\bu,
\bs))\frac{q(\bu\given\bs)}{Q(\bu\given\by,\bH)}\ds+\beta(\by,\bH)
= 0\nonumber\\& &
\quad\quad\quad\quad\quad\quad\quad\quad\quad\quad\quad\quad\quad\quad\quad\quad\quad\quad\quad\quad\quad\quad\quad\quad
\forall\bu\in\RR^M,\forall\by\in\RR^N,\forall\bH\in\cR_H
\label{eq:Lagrange_Q}\end{eqnarray}
\end{definition}
We say that two functions $q$ and $Q$ {\it admit} Lagrange
multipliers if Lagrange multipliers that satisfy
Definition~\ref{def:Lagrangians} exist for them.

To obtain some motivation for~\eqref{eq:Lagrange_q}
and~\eqref{eq:Lagrange_Q}, consider the formal Lagrangian, defined
as
\begin{eqnarray}\label{eq:formal_lag}
\cL(q,Q;\:\Upsilon,\Gamma,\alpha,\beta) &\defined& F(q,Q) + <\Upsilon,\bE(q)> +
<\Gamma,\bC(q)> + \int_{\bs\in\RR^M} \alpha(\bs)\cdot\int_{\bu\in\RR^M}
q(\bu\given\bs)\ d\bu\  d\bs +\nonumber\\&&
+\int_{\by\in\RR^N}\int_{\bH\in\cR_H}
\beta(\by,\bH)\cdot\int_{\bu\in\RR^M} Q(\bu\given\by,\bH)\du\dH\dy
\end{eqnarray}
where $\bE(q)$ and $\bC(q)$ are matrix-valued functionals given by
the left-hand-side of~\eqref{eq:E_q} and~\eqref{eq:C_q}.  Formally
differentiating $\cL(q,Q;\:\Upsilon,\Gamma,\alpha,\beta)$ with respect to
$q(\bu\given\bs)$ (for given $\bu$ and $\bs$)  and comparing with
zero, would render~\eqref{eq:Lagrange_q}.  Similarly,
differentiating with respect to $Q(\bu\given\by,\bH)$ (for given
$\bu$,$\by$ and $\bH$), and comparing with zero, would
render~\eqref{eq:Lagrange_Q}. However, the integrals
in~\eqref{eq:formal_lag} are defined over unbounded sets, making
their rigorous analysis difficult.  We therefore prefer to avoid
the use of~\eqref{eq:formal_lag}, and rely on
Definition~\ref{def:Lagrangians} as the definition for Lagrange
multipliers.

We are now ready for the following lemma,
\begin{lemma}\label{lemma:Lagrange}
Let $\qstar$ and $\Qstar$ be a pair of positive-valued feasible
functions for the problem~\eqref{eq:P}.  Assume once again that
$\Qstar$ is the marginal distribution of $\bU$ given $\by$ and
$\bH$, when the distribution of $\bU$ is determined from the
densities $f(\bs)$ and $\qstar(\bu\given\bs)$.  If $\qstar$ and
$\Qstar$ admit Lagrange multipliers, then they are a solution
(i.e., achieve the global maximum) of~\eqref{eq:P}.
\end{lemma}
A proof of Lemma~\ref{lemma:Lagrange} is provided in
Appendix~\ref{apdx:Proof_Of_Lemma_Lagrange}.  The proof is
basically an application of well-known concepts from convex
optimization theory.   The proof of
Theorem~\ref{theorem:Gaussian_Maximizes} now focuses on showing
that the above defined $q_G$ and $Q_G$ admit Lagrange multipliers.
We begin by providing the expressions for these two densities.

Recall once more that the setting of the problem (see
Definition~\ref{def:setting}) is fixed.  That is, we assume that
$\Sigma_S$, $\Sigma_{S,X}$, $\Sigma_X$ and $\bF$ are given and
fixed.  Also recall that $\bU$ is related to $\bS$ and $\bX$
through $\bU = \bF\bS + \bX$ and that $q_G$ and $Q_G$ correspond
to a choice of $\bX$ that is jointly-Gaussian with $\bS$.

To obtain $q_G$, we observe that since $\bU$ and $\bS$ are
jointly-Gaussian, the conditional distribution of $\bU$ given
$\bS$ is also Gaussian, with mean $\bm_{U \given S}(\bs)$ and
covariance $\Sigma_{U\given S}$ given by (see e.g.~\cite{Kay}),
\begin{eqnarray*}
\bm_{U \given S}(\bs) &=& \E \bU + \Cov(\bU, \bS)
\cdot\Sigma_S^{-1}\cdot(\bs - \E\bS) \\
\Sigma_{U \given S} &=& \Cov(\bU) - \Cov(\bU, \bS)
\cdot\Sigma_S^{-1}\cdot\Cov(\bS, \bU)
\end{eqnarray*}
Note that by our second regularity assumption (above), that the
covariance of $(\bU,\bS)$ is nonsingular (positive definite), it
follows that $\Sigma_{U \given S}$ is also nonsingular\footnote{
To see this, assume by contradiction that $\bv\Sigma_{U \given
S}\bv^T = 0$ for some nonzero row vector $\bv$.  Thus, with
probability 1 we would have $\bv \cdot \bU = \bv \cdot \bm_{U
\given S}(\bS)$, and therefore, using~\eqref{eq:22}, $[\bv, -\bv
J]\cdot[\bU^T,\bS^T]^T = 0$. This would imply that $\Cov(\bU,\bS)$
is singular.}.

Using $\bU = \bF \bS + \bX$ and $\E \bU = \E \bS = \bzr$, we
obtain,
\begin{eqnarray}
\bm_{U \given S}(\bs) &=& J \bs,\quad\textrm{where}\quad J \defined (\bF\Sigma_S + \Sigma_{S,X}^T)\Sigma_S^{-1} \label{eq:22}\\
 \Sigma_{U \given S}  &=&
(\bF\Sigma_S\bF^T + \bF\Sigma_{S,X} + \Sigma_{S,X}^T\bF^T +
\Sigma_X) - (\bF\Sigma_S +
\Sigma_{S,X}^T)\Sigma_S^{-1}(\bF\Sigma_S + \Sigma_{S,X}^T)^T
\label{eq:23}
\end{eqnarray}
Observe that $J$ and $\Sigma_{U \given S}$ are fixed matrix
functions of the matrices $\Sigma_X$, $\Sigma_S$, $\Sigma_{S,X}$
and $\bF$ that constitute the problem setting. Hence,
\begin{eqnarray}\label{eq:q_G}
q_G(\bu\given\bs) = \frac{1}{\sqrt{\det(2\pi \Sigma_{U\given S}
)}}\exp(-\frac{1}{2}(\bu-J\bs)^T \Sigma_{U\given
S}^{-1}(\bu-J\bs)) \quad \bs \in \RR^M, \bu \in \RR^M,
\end{eqnarray}

To obtain $Q_G$, we observe that for fixed $\bH$, the distribution
of $\bU$ given $\bY$ is also Gaussian.
\begin{eqnarray*}
\bm_{U \given Y,H}(\by,\bH) &=& \E [\bU\given H = \bH] + \Cov(\bU,
\bY \given H = \bH)
\cdot\Cov(Y\given H = \bH)^{-1}\cdot(\by - \E[\by \given H = \bH]) \label{eq:24}\\
\Sigma_{U \given Y, H}(\bH) &=& \Cov(\bU \given H = \bH) -
\Cov(\bU, \bY \given H = \bH) \cdot\Cov(Y\given H =
\bH)^{-1}\cdot\Cov(\bY, \bU \given H = \bH)
\end{eqnarray*}
We now claim that $\Sigma_{U \given Y, H}(\bH)$ is also
nonsingular. This will be shown by proving that $\Sigma_{U,Y |
H}(\bH)$ is positive definite, i.e.
\begin{equation}
\label{eq:need_to_prove} (\balpha^T, \bbeta^T) \Sigma_{U,Y |
H}(\bH) \left( \begin{array}{c} \balpha \\ \bbeta \\ \end{array}
\right) = {\rm E} \left\{ \left( \balpha^T \bU + \bbeta^T \bY
\right)^2 \given H = \bH \right\} > 0 \quad \forall (\balpha ,
\bbeta) \neq \bzr
\end{equation}
Now, by~(\ref{eq:FadingDirtyPaper}) and~(\ref{eq:F}),
$$
\bY = H (-\bF + \bI) \bS + H \bU + \bZ
$$
By our second regularity assumption, the covariance of $(\bU,\bS)$
is nonsingular. It follows that $\Sigma_U$ is positive definite.
We thus conclude that~(\ref{eq:need_to_prove}) holds for
$\beta=0$. If, on the other hand $\beta \neq \bzr$, then
$$
{\rm E} \left\{ \left( \balpha^T \bU + \bbeta^T \bY \right)^2
\given H = \bH \right\} = {\rm E} \left\{ \left( \balpha^T \bU +
\bbeta^T \left( H (-\bF + \bI) \bS + H \bU \right) \right)^2
\given H = \bH \right\} + {\rm E} \left\{ \left( \bbeta^T \bZ
\right)^2 \right\}
> 0
$$
since $Z$ is independent of $\bX$, $\bS$ and $H$, and its
covariance, $\Sigma_Z$, is nonsingular. This proves our claim.

Using similar arguments as in the above development of $q_G$, we
obtain
\begin{eqnarray*}
\bm_{U \given Y,H}(\by,\bH) = K(\bH)\by
\end{eqnarray*}
where,
\begin{eqnarray*}
K(\bH) = \left[(\bF\Sigma_S + \bF\Sigma_{S,X} + \Sigma_{S,X}^T +
\Sigma_X)\bH^T \right]\left[ \bH(\Sigma_S + \Sigma_X +
\Sigma_{S,X} + \Sigma_{S,X}^T)\bH^T + \Sigma_Z \right]^{-1}
\end{eqnarray*}
and,
\begin{eqnarray}
&&\Sigma_{U \given Y, H}(\bH) = (\bF\Sigma_S\bF^T +
\bF\Sigma_{S,X} + \Sigma_{S,X}^T\bF^T + \Sigma_X) -\nonumber
\\&&\quad
\left[(\bF\Sigma_S + \bF\Sigma_{S,X} + \Sigma_{S,X}^T +
\Sigma_X)\bH^T \right] \left[\bH(\Sigma_S + \Sigma_X +
\Sigma_{S,X} + \Sigma_{S,X}^T)\bH^T + \Sigma_Z \right]^{-1}
\times\nonumber\\&&\quad\quad\quad\quad\quad\quad\quad\quad\quad\left[(\bF\Sigma_S
+ \bF\Sigma_{S,X} + \Sigma_{S,X}^T + \Sigma_X)\bH^T \right]^T
\label{eq:26}
\end{eqnarray}
Observe that $K(\bH)$ and $\Sigma_{U \given Y,H}(\bH)$ are fixed
matrix functions of the matrices that constitute the problem
setting, and of $\bH$. Hence,
\begin{eqnarray}\label{eq:Q_G}
Q_G(\bu\given\by, \bH) &=& \frac{1}{\sqrt{\det(2\pi
\Sigma_{U\given Y, H}(\bH) )}}\exp(-\frac{1}{2}(\bu-K(\bH)\by)^T
\Sigma_{U\given Y, H}(\bH)^{-1}(\bu-K(\bH)\by)) \nonumber
\\&&\quad\quad\quad\quad\quad\quad\quad\quad\quad\quad\quad\quad\quad\quad\quad\quad  \by \in \RR^N, \bH \in \cR_H,
\bu \in \RR^M
\end{eqnarray}

We observe that $q_G(\bu\given\bs)$ is positive-valued for all
$\bu \in \RR^M$ and $\bs \in \RR^M$. Similarly, $Q_G(\bu\given\by,
\bH)$ is positive-valued for all $\bu \in \RR^M$, $\by \in \RR^N$
and $\bH \in \cR_H$, where $\cR_H$ is the support region of $H$.
Therefore, they satisfy this condition of
Lemma~\ref{lemma:Lagrange}. The conditions of
Lemma~\ref{lemma:Lagrange} also require that $Q_G$ be the marginal
distribution of $\bU$ given $\by$ and $\bH$, when the distribution
of $\bU$ is determined from the densities $f(\bs)$ and
$q_G(\bu\given\bs)$. This is satisfied by definition.

We proceed by showing that the two functions $q_G$ and $Q_G$ admit
Lagrange multipliers. Finding a Lagrange multiplier $\beta(\by,
\bH)$ to satisfy~\eqref{eq:Lagrange_Q} is easy.  As in the
discussion following~\eqref{eq:7}, we have
\begin{eqnarray*} \int_{\bs\in \RR^M}f(\bs)f(\by,\bH\given \bs,
\bx = \bof(\bu,
\bs))\frac{q_G(\bu\given\bs)}{Q_G(\bu\given\by,\bH)}\ds =
q_G(\by,\bH) \quad\quad
\forall\bu\in\RR^M,\forall\by\in\RR^N,\forall\bH\in\cR_H
\end{eqnarray*}
Thus, defining $\beta(\by, \bH) =
-q_G(\by,\bH)$,~\eqref{eq:Lagrange_Q} is satisfied.

We now turn our attention to the other Lagrange multipliers and
to~\eqref{eq:Lagrange_q}. Let $\bu$ and $\bs$ be fixed and let
$\bx
\defined \bof(\bu, \bs)$.  Simple manipulations of~\eqref{eq:Lagrange_q}
lead to,
\begin{eqnarray*}
&&\int_{\by\in \RR^N}\int_{\bH\in\cR_H}f(\by,\bH\given \bs,
\bx)\log{Q_G(\bu\given\by,\bH)}\dH \dy -
\\&&\quad\quad\quad\quad
-\int_{\by\in \RR^N}\int_{\bH\in\cR_H}f(\by,\bH\given \bs, \bx)\dH
\dy \cdot\left[\log{q_G(\bu\given\bs)} + 1\right]+\\&
&\quad\quad\quad\quad +
<\Upsilon,\bx\cdot\bx^T>+<\Gamma,\bs\cdot\bx^T>+\frac{\alpha(\bs)}{f(\bs)}
= 0
\end{eqnarray*}
We continue,
\begin{eqnarray}
&&\int_{\by\in \RR^N}\int_{\bH\in\cR_H}f(\by,\bH\given \bs,
\bx)\log{Q_G(\bu\given\by,\bH)}\dH \dy - \log{q_G(\bu\given\bs)} +
<\Upsilon,\bx\cdot\bx^T>+<\Gamma,\bs\cdot\bx^T>\nonumber\\&
&\quad\quad\quad\quad+\left[\frac{\alpha(\bs)}{f(\bs)}-1\right] =
0\label{eq:9}
\end{eqnarray}
We begin by examining the first element in the above sum.  This
element is equal to,
\begin{eqnarray}
&&\E_{Y,H} \left[\log{Q_G(\bu\given\bY,H)}\given \bX = \bx, \bS =
\bs\right]= \nonumber\\&&\quad\quad = -\frac{1}{2}\E_{Y,H}
\left[\log\det(2\pi \Sigma_{U\given Y, H}(H) )\given \bx,
\bs\right] \nonumber\\&&\quad\quad \: \: \: \:
-\frac{1}{2}\E_{Y,H}\left[(\bu-K(H)\bY)^T \Sigma_{U\given Y,
H}(H)^{-1}(\bu-K(H)\bY)\given \bx,
\bs\right]\nonumber\\
&&\quad\quad = -\frac{1}{2}\E_{H} \left[\log\det(2\pi
\Sigma_{U\given Y, H}(H) )\right] \nonumber\\&&\quad\quad \: \: \:
\: -\frac{1}{2}\E_{H}\left\{\E_{Y}\left[(\bu-K(H)\bY)^T
\Sigma_{U\given Y, H}(H)^{-1}(\bu-K(H)\bY)\given \bx, \bs, H
\right]\right\}\label{eq:8}
\end{eqnarray}
We now focus on the contents of the braces.  We use $\bu = \bF\bs
+ \bx$, $\bY = \bH(\bx + \bs) + \bZ$ to obtain,
\begin{eqnarray*}
&&\E_{Y}\left[(\bu-K(\bH)\bY)^T \Sigma_{U\given Y,
H}(\bH)^{-1}(\bu-K(\bH)\bY)\given \bx, \bs, \bH \right] =\\&&\quad
\bx^T\left[(I-K(\bH)\bH)^T\Sigma_{U\given Y,
H}(\bH)^{-1}(I-K(\bH)\bH)\right]\bx
+\\&&\quad+\bs^T\left[(\bF-K(\bH)\bH)^T\Sigma_{U\given Y,
H}(\bH)^{-1}(\bF-K(\bH)\bH)\right]\bs +\\&&\quad
2\bs^T\left[(\bF-K(\bH)\bH)^T\Sigma_{U\given Y,
H}(\bH)^{-1}(I-K(\bH)\bH)\right]\bx +\\&&\quad+
\tr\left[K(\bH)^T\Sigma_{U\given Y, H}(\bH)^{-1}K(\bH) + \Sigma_Z
\right]
\end{eqnarray*}
Thus, we can rewrite~\eqref{eq:8} as,
\begin{eqnarray}
\bx^TA\bx +\bs^TB\bs + \bs^TC\bx + D = <A, \bx\cdot\bx^T> +
<B,\bs\cdot\bs^T> + <C,\bs\cdot\bx^T> + D \label{eq:12}
\end{eqnarray}
where,
\begin{eqnarray}
A &=& -\frac{1}{2}E_{H}\left[(I-K(H)H)^T\Sigma_{U\given Y,
H}(H)^{-1}(I-K(H)H)\right]\label{eq:27}\\B &=&
-\frac{1}{2}E_{H}\left[(\bF-K(H)H)^T\Sigma_{U\given Y,
H}(H)^{-1}(\bF-K(H)H)\right]\label{eq:28}\\ C &=&
-E_{H}\left[(\bF-K(H)H)^T\Sigma_{U\given Y,
H}(H)^{-1}(I-K(H)H)\right]\label{eq:29}\\D &=& -\frac{1}{2}\E_{H}
\left[\log\det(2\pi \Sigma_{U\given Y, H}(H) )\right]
-\frac{1}{2}E_{H} \left\{\tr\left[K(H)^T\Sigma_{U\given Y,
H}(H)^{-1}K(H) + \Sigma_Z \right]\right\}\label{eq:30}
\end{eqnarray}
By the conditions of Theorem~\ref{theorem:Gaussian_Maximizes}, the
above expectations exist and are finite.  Turning to the second
element of the sum in~\eqref{eq:9} we obtain, using~\eqref{eq:q_G}
\begin{eqnarray}\label{eq:11}
-\log q_G(\bu\given\bs) = \frac{1}{2}\log \det(2\pi
\Sigma_{U\given S} ) + \frac{1}{2}(\bu-J\bs)^T \Sigma_{U\given
S}^{-1}(\bu-J\bs)
\end{eqnarray}
Applying a similar development to that of~\eqref{eq:8}, we can
rewrite~\eqref{eq:11} as,
\begin{eqnarray}\label{eq:13}
<\hA, \bx\cdot\bx^T> + <\hB,\bs\cdot\bs^T> + <\hC,\bs\cdot\bx^T> +
\hD
\end{eqnarray}
where
\begin{eqnarray*}
\hA &=& \frac{1}{2}\Sigma_{U\given S}^{-1}\\\hB &=&
\frac{1}{2}(\bF-J)^T\Sigma_{U\given S}^{-1}(\bF-J)\\ \hC &=&
(\bF-J)^T\Sigma_{U\given S}^{-1}\\\hD &=& \frac{1}{2}
\log\det(2\pi \Sigma_{U\given S} )
\end{eqnarray*}
Using~\eqref{eq:12} and~\eqref{eq:13}, we can rewrite~\eqref{eq:9}
as,
\begin{eqnarray*}
&&<A+\hA+\Upsilon, \bx\cdot\bx^T> + <B+\hB,\bs\cdot\bs^T> +
<C+\hC+\Gamma,\bs\cdot\bx^T> + D+\hD
+\left[\frac{\alpha(\bs)}{f(\bs)}-1\right] = 0
\end{eqnarray*}
Finally, we may select our Lagrange multipliers
for~\eqref{eq:Lagrange_q} as follows, completing the proof of
Theorem~\ref{theorem:Gaussian_Maximizes}.
\begin{eqnarray*}
\Upsilon = -(A+\hA),\quad \Gamma = -(C+\hC),\quad \alpha(\bs) =
f(\bs)\left[1 - D - \hD -<B+\hB,\bs\cdot\bs^T>\right]
\end{eqnarray*}
\end{proof}

Note that with linear-assignment, when $\bX$ and $\bS$ are
jointly-Gaussian, the achievable rate $I(\bU;\bY,H) - I(\bU;\bS)$
is a function of the setting (as defined in
Definition~\ref{def:setting}).  The expression for the achievable
rate can be computed as follows,
\begin{eqnarray}\label{eq:Gauss_LA_Achievable}
I(\bU;\bY,H) - I(\bU;\bS) = h(\bU\given\bS) - h(\bU\given\bY,H) =
\frac{1}{2}\log\det\Sigma_{U\given S} -
\frac{1}{2}E_H\left[\log\det\Sigma_{U\given Y,H}(H)\right]
\end{eqnarray}
The last equation is obtained from the following discussion.  For
fixed $\bs$, the marginal distribution of $\bU$ given $\bS=\bs$ is
zero-mean Gaussian distributed with variance $\Sigma_{U \given S}$
(which is given by~\eqref{eq:23} and is independent of $\bs$). For
fixed $\by$ and $\bH$, the marginal distribution of $\bU$ given
$\bY=\by$ and $H = \bH$ is zero-mean Gaussian distributed with
variance $\Sigma_{U \given Y,H}(\bH)$ (which is given
by~\eqref{eq:26} and is independent of $\by$ but dependent on
$\bH$).

Note that the achievability proof of Gel'fand and
Pinsker~\cite{gelpinsk:80}, that states that we may indeed achieve
the rate $I(\bU;\bY;H) - I(\bU;\bS)$ assumes that the random
variables involved are discrete-valued.
In Appendix~\ref{apdx:Achievability_FqGQG} we use quantization arguments to prove that
$F(q_G,Q_G)$, defined using~\eqref{eq:F_q_Q_defined} (which
assumes {\it continuous} random variables), is indeed achievable.

\section{The Linear-Assignment Fading-Paper (LAFP) Achievable Region}\label{sec:Fading_Paper_Achievable}
\subsection{Definition}\label{subsec:LAFP_Def}
In Sec.~\ref{sec:Dirty_Paper_Achievable} we described how
dirty-paper transmission methods can be used to construct an
algorithm for transmission over the non-fading MIMO-BC channel.
The same approach can be used to construct an algorithm for
transmission over the fading MIMO-BC channel, using the
linear-assignment fading-paper transmission methods of
Sec.~\ref{sec:Fading_Paper}.

In our approach, we rely on
Theorem~\ref{theorem:Gaussian_Maximizes} and confine our attention
to Gaussian distributions for the signals $\{\bX_l\}_{l=1}^L$,
defined as in Sec.~\ref{sec:Dirty_Paper_Achievable}.
Our choice is greedy in the sense that we seek to maximize the rate to each user
individually, while a global perspective could possibly prescribe
a different choice. However, a similar choice in the definition of
the dirty-paper achievable region was eventually proven to
coincide with the global optimum as well.  We refer to the
convex-hull of the union of rate regions that are achievable using
this approach, as the {\it linear-assignment fading-paper} (LAFP)
achievable region.

The analysis of Weingarten~\etal~\cite{HananYossiShlomo} does not
apply to the fading setting.  Furthermore, linear-assignments have
not been proven to exhaust the capacity of the fading-paper
channel.  Thus, unlike the dirty-paper achievable region of
Sec.~\ref{sec:Dirty_Paper_Achievable}, the LAFP achievable region
is not guaranteed to be optimal.

The determination of the dirty paper achievable region of
Sec.~\ref{sec:Dirty_Paper_Achievable} involves determining the
covariance matrices $\Sigma_X^{(l)}$ for the various signals
$\bX_l$ (see e.g.~\cite{Caire_Shamai_MIMO} and
\cite{Goldsmith_MIMO_BC}).  However, each signal $\bX_l$ is
assumed to be independent of the interference $\bS_l \defined
\sum_{i < l}\bX_i$, and Gaussian.  In our above definition of the
LAFP, we have not restricted ourselves to signals
$\{\bX_l\}_{l=1}^L$ that are independent of their respective
interferences $\{\bS_l\}_{l=1}^L$. Thus, in addition to
determining $\Sigma_{X}^{(l)}$, it would appear that we must
determine the covariance $\Sigma_{X,S}^{(l)}$, between $\bX_l$ and
$\bS_l$ as well.

However, the following theorem proves that we may indeed confine
ourselves to $\Sigma_{X,S}^{(l)} = \bzr$, without loss of
optimality.
\begin{theorem}\label{theorem:Independent_Maximizes}
The LAFP achievable region is exhausted by a choice of random
variables $\{\bX_l\}_{l=1}^L$ for the various users that are
independent of their respective interferences
$\{\bS_l\}_{l=1}^{L}$
\end{theorem}
The proof of this theorem is provided in
Appendix~\ref{apdx:Proof_Independent_Maximizes}.

Note that in this theorem we do {\it not} claim that for the given
fading-paper problem observed by user $l$, selecting $\bX_l$ to be
independent of $\bS_l$ incurs no loss of optimality.  Rather, the
proof involves replacing an entire given set of signals
$\bX_1,...,\bX_L$, which may not be independent (corresponding to
some set of achievable rates on the LAFP achievable region) with a
new set $\hbX_1,...,\hbX_L$ that are independent, without
sacrificing the rates of the individual users.  In the resulting
set, user $l$'s signal $\hbX_l$ is indeed independent of $\hbS_l =
\sum_{i<l}\hbX_i$.  However, the independence was achieved also by
altering the fading-paper problem this user faces.
\subsection{Comparison with Dirty-Paper Transmission}
\label{sec:Comparison Dirty Paper}

So far, we have focused on similarities between the dirty-paper
transmission over a fixed MIMO-BC and LAFP transmission over a
fading MIMO-BC channel.  Both approaches use linear strategies,
both employ independently distributed Gaussian random variables to
construct their signals to the receivers.

However, the two methods differ in two important ways.
\begin{enumerate}
\item The choice of the constant matrix $\bF$ in dirty-paper
transmission is based on the fixed channel matrix $\bH$. With
fading-paper, only the statistics of $\bH$ are known and thus
$\bF$ must be selected differently.
\item The fading-paper receiver accounts for a channel
fade $\bH$ that fluctuates from one time instance to another.  The
dirty-paper receiver assumes that $\bH$ is fixed.
More precisely, the dirty paper decoder seeks a codeword that is jointly typical with
$\by$, while the fading paper decoder seeks a codeword that is jointly typical with both $\by$ and $\bH$.
\end{enumerate}
Despite these two shortcomings, dirty-paper transmission can still
be applied to a fading-paper channel by simply assuming that $\bH$
is fixed at its average, and treating its fluctuations as noise.
For a fading paper transmission strategy to be interesting, we
must demonstrate that its performance surpasses that of
dirty-paper transmission.

An evaluation of the dirty-paper achievable region (i.e., when the
transmitter and receiver assume that the channel is fixed at its
expected value $\E H$) over the fading MIMO-BC scheme is
difficult. This is because of the operation of the decoder, which
uses a mismatched model of the channel. However, we may obtain an
outer bound on the dirty-paper achievable region if we replace the
receiver with an optimal LAFP receiver that uses the channel
information available to it (unlike the standard dirty-paper
receiver). In this case, the achievable rate may be obtained
from~\eqref{eq:Gauss_LA_Achievable}. With the dirty-paper
achievable region, however, the matrices $\bF$
(for each instance of $\Sigma_X$, $\Sigma_S$ and $\Sigma_Z$ for the user) are not the optimal fading paper
matrices, but rather are computed using~\eqref{eq:F_DPC}, under
the assumption of a fixed channel matrix, equal to $\E H$. Under
these conditions, the approach differs from LAFP only in the way
the matrix $\bF$ is selected.

We let $\bF_{DPC}(\bH)$ denote the choice of $\bF$ with
dirty-paper transmission over a channel whose fixed channel matrix
is $\bH$. That is, $\bF_{DPC}(\bH)$ is a matrix function of $\bH$,
given by the right hand side of~\eqref{eq:F_DPC}
(for brevity of notation, we neglect the reliance of $F_{DPC}(\cdot)$ on $\Sigma_X$ and
$\Sigma_Z$). With this notation, the choice of $\bF$ that is used
in the above-mentioned dirty paper like transmission strategy is
$\bF_{DPC}(\E H)$.

Evaluating the LAFP region involves determining the union of the regions obtained for all
matrices $\bF$.  Equivalently, it involves
maximizing~(\ref{eq:Gauss_LA_Achievable}) over $\bF$ (e.g. using a grid search) given the
covariances of $X$ and $S$ (note that by Theorem~\ref{theorem:Independent_Maximizes} we set $\Sigma_{S,X}=0$).
However, we obtained an {\it inner} bound by restricting our attention, for
each $\Sigma_X$ and $\Sigma_Z$  to the set
\begin{eqnarray} \label{eq:F_LAFP}
\cF \defined \left\{\:\bF_{DPC}(\bH)\::\: \bH \in \cR_H\:\right\}
\end{eqnarray}

\begin{figure}[htp]
               \begin{center}
               \leavevmode
               \centerline{\resizebox{0.6\textwidth}{!}{\includegraphics{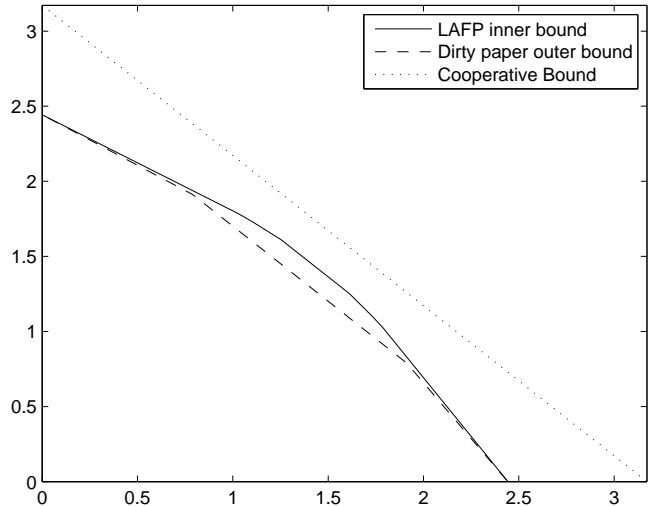}}}
               \vspace*{-0.2in}\caption{Comparison between an inner bound on the LAFP achievable region and an outer bound on the dirty paper achievable region.}
               \label{fig:Comparison}
               \end{center}
\end{figure}
Fig.~\ref{fig:Comparison} presents a numerical example where the
two approaches are compared.  In this example, there are two users
(receivers). The transmitter has two antennas ($M = 2$) and the
receivers have one antenna each $(N_1 = N_2 = 1)$.  The power
constraint is $P_{TOT} = 10$.   The distributions of the channel
matrices are given by,
\begin{eqnarray*}
H^{(1)} = \cases{[1, 0.4] & with probability 1/2 \cr [1, 3] & with
probability 1/2}\quad H^{(2)} = \cases{[0.4, 1] & with probability
1/2 \cr [3, 1] & with probability 1/2}
\end{eqnarray*}
The noise variance at each receiver is $1$.

The achievable regions in both cases (i.e. LAFP and dirty-paper)
were found by first applying a grid search for the matrices
$\Sigma_X^{(1)}$ and $\Sigma_X^{(2)}$.  In line with
Theorem~\ref{theorem:Independent_Maximizes}, we assumed without
loss of optimality that the two signals $\bX^{(1)}$ and
$\bX^{(2)}$ are independent.

For each such pair $\Sigma_X^{(1)}$ and $\Sigma_X^{(2)}$, the
matrix $\bF$ for user 2 was computed as described above.  That is,
for the LAFP achievable region, $\bF$ was found by maximizing the
achievable rate of user 2 over the set $\cF$ (which is a function
of the user's covariance matrix\footnote{In the general case,
where there are more than two users, $\cF$ is also a function of
$\sum_{l > 2}\Sigma_X^{(l)}$, the unknown interference from
subsequent users, which must be accounted for in the effective
noise as explained in Appendix~\ref{apdx:LAFP_L_GT_2}.}
$\Sigma_X^{(2)}$).  For the dirty-paper achievable region,
$\bF_{DPC}(\E H)$ was used.

With both schemes, for fixed matrices $\Sigma_X^{(1)}$,
$\Sigma_X^{(2)}$ and $\bF$, the achievable rates $R_1$ and $R_2$
for the two users were computed as follows.  $R_1$ was obtained
using the following expression (recall that user 1's observed
signal $Y^{(1)}$ is scalar in this example),
\begin{eqnarray*}
R_1 = \frac{1}{2}E_{H^{(1)}} \log \left(1 +
\frac{H^{(1)}\Sigma_X^{(1)}H^{(1)T}}{H^{(1)}\Sigma_X^{(2)}H^{(1)T}
+ 1}\right)
\end{eqnarray*}
$R_2$ is given by the right hand side
of~\eqref{eq:Gauss_LA_Achievable}. Since we have assumed
$\bX^{(1)}$ and $\bX^{(2)}$ to be independent, the expressions for
$\Sigma_{U\given S}$ and $\Sigma_{U \given Y,H}(\bH)$ (which
appear in~\eqref{eq:Gauss_LA_Achievable}) are simple\footnote{In
the context of our discussion, $\bS = \bX^{(1)}$, $\bX =
\bX^{(2)}$ and $\bZ$ has covariance $\Sigma_Z = 1$. $\bU = \bF \bS
+ \bX$, as usual.}.  That is, $\Sigma_{U\given S} =
\Sigma_X^{(2)}$ and $\Sigma_{U \given Y,H}(\bH)$ is obtained
from~\eqref{eq:26} by setting $\Sigma_{S,X}$ to zero.

The maximal sum-rate on the dirty paper outer bound was 2.7 bits
per channel use, while the maximum sum-rate on the LAFP inner
bound was 2.86.  This achievable rate was obtained by selecting,
\begin{eqnarray*}
\Sigma_X^{(1)} = \left[\begin{array}{cc}
  1 &  2-\epsilon \\
  2-\epsilon & 4
\end{array}\right],\quad
\Sigma_X^{(2)} =\left[\begin{array}{cc}
  4.5 &  -1.5+\epsilon \\
  -1.5+\epsilon & 0.5
\end{array}\right],\quad
\bF = \left[\begin{array}{cc}
  1.0909  &  0.3636\\
   -0.3636 &  -0.1212
\end{array}\right]
\end{eqnarray*}
where $0<\epsilon\rightarrow 0$ such that $\Sigma_X^{(1)}$ and $\Sigma_X^{(2)}$ are positive definite.
Thus, a simple approach, which uses knowledge of the
channel distribution at the transmitter, was able to produce {\em at least}
a 6\% increase in throughput.

Although we have not established the optimality of the LAFP
achievable region, we can obtain an idea of how far we are from
the optimum using a cooperative upper bound on the achievable
sum-capacity (i.e., the maximum achievable sum rate to all users),
as suggested by Sato~\cite{Sato}.  The use of such a bound in the
context of the (non-fading) MIMO-BC channel was first suggested by
Caire and Shamai~\cite{Caire_Shamai_MIMO}. Computation of
cooperative upper-bounds for the above fading MIMO-BC example is
discussed in Appendix~\ref{apdx:Sato}.  We obtained a bound of
3.17 on the maximum achievable sum-rate. Thus, in terms of the
sum-rate, LAFP is capable of transmission at rates that are 10\% below the
optimum.

In Appendix~\ref{apdx:LAFP_L_GT_2} we will discuss the computation
of the LAFP achievable region with more than two users.

\section{Conclusion} \label{sec:Conclusion}
\subsection{Suggestions for Further Research} \label{subsec:Further_Research}
\begin{enumerate}
\item {\bf Heuristic methods for computing $\bF$.}
Expression~\eqref{eq:F_LAFP}, with which we computed the matrix
$\bF$ for the LAFP region in Sec.~\ref{sec:Comparison Dirty
Paper}, was developed heuristically.  A different expression could
possibly produce a substantially larger achievable region. One
option would be to search for $\bF$ along a fine grid (as noted in
Sec.~\ref{sec:Comparison Dirty Paper}). An alternative option
would be to apply a gradient ascent method, using $\bF$ as defined
in~\eqref{eq:F_LAFP} as a starting point.
\item {\bf A wider range of strategies.}  The confinement to
linear assignments as defined in Sec.~\ref{sec:Fading_Paper} is in
no way known to be optimal.  Dupuis~\etal~\cite{Dupuis_Blahut_GP}
suggested an algorithm that is based on the concepts of the
Blahut-Arimoto algorithm, that can theoretically be used to
evaluate the capacity of a general side-information channel (of
which the fading paper channel is an instance).  In practice,
applying the algorithm requires evaluations over a set of {\it
strategies} which is impossibly large.  However, applying the
algorithm over any subset of these strategies produces an
achievable rate.
This achievable rate may further narrow the gap to the cooperative
upper-bound (as discussed in Sec.~\ref{sec:Comparison Dirty
Paper}).
\end{enumerate}
\subsection{Concluding Remarks}
The problem of transmitting over fading MIMO-BC channels is of
great practical interest.  In this paper we  presented an
achievable region for this channel that relies on fading-paper
transmission strategies.  Our main contribution is
Theorem~\ref{theorem:Gaussian_Maximizes}, which proves that a
Gaussian distribution achieves the linear-assignment capacity.  We
believe that the approach we developed in the proof of that
theorem, which employs convex-analysis methods, could be useful in
further analysis of this channel.

In Sec.~\ref{sec:Comparison Dirty Paper} we have shown that a
simple approach, which makes use of the channel distribution
information available to the transmitter, easily produces a gain
over dirty-paper transmission.  Further research (perhaps in the
lines of Sec.~\ref{subsec:Further_Research}) could produce further
performance gains.
\appendices
\section{The Optimal Achievable Rate with Zero Channel State Information at the Transmitter}\label{apdx:TDMA}
Consider a broadcast channel where all the receivers have the same
number of antennas.  We wish to show that capacity in this case is
achieved by time-sharing among the users.

A channel model that assumes zero knowledge of the channel fade to
each of the users, effectively assumes that all channels are the
same.  The signals at the different receivers are equivalent in
their statistical properties, and thus each receiver is capable,
beside decoding its own signal, of decoding all the messages to
the other users as well.  Thus, the sum-rate of this system is
upper-bounded by the single-user rate of each of the users.  Such
a capacity region is exhausted by time-sharing.
\section{The Optimal Matrix $\bF$ in the Achievability Proof for Dirty-Paper}\label{apdx:Proof_F}
In this appendix we prove the optimality of $\bF$ as defined
by~\eqref{eq:F_DPC}.  We let $\bU$ and $\bX$ be defined as in the
discussion preceding~\eqref{eq:F_DPC}.  The achievable rate with
this choice is given by $I(\bU;\bY) - I(\bU,\bS)$
(see~\eqref{eq:GelPinsk}). We now seek to prove that this rate
coincides with the capacity of the corresponding no-interference
channel defined by~\eqref{eq:No_Interference}. Our proof follows
in the lines of a similar proof by Cohen and
Lapidoth~\cite{Caire_Shamai_MIMO} for the scalar dirty-paper
channel.

To obtain our result, we prove a stronger result.  We prove that
for any choice of $\Sigma_X$, letting $\bF$ be given
by~\eqref{eq:F_DPC}, we obtain that the achievable rate coincides
with the achievable rate $I(\bX;\hbY)$ for the no-interference
channel~\eqref{eq:No_Interference}.

Our objective is to show that the achievable rate $I(\bU;\bY) -
I(\bU,\bS)$, with this choice of $\bF$, coincides with the
achievable rate of the no-interference channel when the input
$\bX$ is distributed as $\cN(\bzr, \Sigma_X)$.

Let $\hbX = \bW \hbY$ be the linear minimum mean-square error
(LMMSE) estimate for $\bX$ given $\hbY$.  $\bW$ is obtained
by~\cite{Kay},
\begin{eqnarray}\label{eq:W}
\bW = \Cov(\bX,\hbY)\Cov(\hbY)^{-1} =
\Sigma_X\bH^T(\bH\Sigma_X\bH^T + \Sigma_Z)^{-1}
\end{eqnarray}
By definition of the LMMSE estimate, the error $\bE
\defined \bX - \hbX$ is uncorrelated with $\hbY$.  Since $\bE$ and
$\hbY$ are jointly-Gaussian, they are also independent.  $\bS$ is
independent of both, and thus $\bE$ is independent of $\bY = \hbY
+ \bH\bS$.

Examining $I(\bU;\bY) - I(\bU,\bS)$, we have
\begin{eqnarray} \label{eq:IUY_IUS}
I(\bU;\bY) - I(\bU,\bS) = h(\bU\given\bS) - h(\bU\given\bY)
\end{eqnarray}
We now examine both elements of the difference on the right hand
side of the above.
\begin{eqnarray}\label{eq:hUS}
h(\bU\given\bS) = h(\bF\bS + \bX \given \bS) = h(\bX)
\end{eqnarray}
where the last equation is obtained by the fact that $\bS$ and
$\bX$ are independent.
\begin{eqnarray}\label{eq:hUY}
h(\bU\given\bY) &=& h(\bF\bS + \bX \given \bY) \refeq{a}
h(\bW\bH\bS + \bX \given \bY) \nonumber\\&=& h(\bW\bH\bS + \bX -
\bW\bY\given \bY) = h(\bW\bH\bS + \bX - \bW(\bH\bS + \bH\bX +
\bZ)\given \bY) \nonumber\\&=& h(\bX - \bW(\bH\bX + \bZ)\given
\bY) = h(\bX - \hbX\given \bY) = h(\bE\given \bY) \refeq{b} h(\bE)
\refeq{c} h(\bE\given\hbY)
\nonumber\\&=&
h(\bX - \bW\hbY \given\hbY)
=
h(\bX \given\hbY)
\end{eqnarray}
Equality (a) is obtained from the observation that the right hand
side of~\eqref{eq:F_DPC} equals $\bW\cdot\bH$ where $\bW$ is given
by~\eqref{eq:W}.  Equalities (b) and (c) are obtained from the
fact that $\bE$ is independent of $\hbY$ and $\bY$.  Finally,
combining \eqref{eq:IUY_IUS},~\eqref{eq:hUS} and~\eqref{eq:hUY} we
obtain our desired result,
\begin{eqnarray*}
I(\bU;\bY) - I(\bU,\bS) = h(\bX) - h(\bX \given \hbY) =
I(\bX;\hbY)
\end{eqnarray*}
\QEDopen

\section{Proof of Lemma~\ref{lemma:Lagrange}} \label{apdx:Proof_Of_Lemma_Lagrange}
Let $q$ and $Q$ be a pair of feasible functions for~\eqref{eq:P}.
We will now show that $F(q,Q) \leq F(\qstar,\Qstar)$.
\begin{eqnarray}\label{eq:1}
F(q,Q) - F(\qstar,\Qstar) &=&
\int_{\bs\in\RR^M}\int_{\bu\in\RR^M}\int_{\by\in
\RR^N}\int_{\bH\in\cR_H}f(\bs)f(\by,\bH\given \bs, \bx = \bof(\bu,
\bs))\cdot\nonumber\\&& \cdot
\left[q(\bu\given\bs)\log\frac{Q(\bu\given\by,\bH)}{q(\bu\given\bs)}-\qstar(\bu\given\bs)
\log\frac{\Qstar(\bu\given\by,\bH)}{\qstar(\bu\given\bs)}\right]
\dH\dy\du\ds
\end{eqnarray}
Let $l(x,y) \defined x\cdot\log(y/x)$.  This function is
jointly-concave in its arguments.  By the gradient
inequality~\cite{boyd}[Chapter 3, Section 3.1.3] for
concave functions, we have for arbitrary $x,y \in \RR_+$ and
$\xstar,\ystar \in \RR_{++}$,
\begin{eqnarray*}
l(x,y) - l(\xstar,\ystar) \leq l_x(\xstar,\ystar)\cdot(x - \xstar)
+ l_y(\xstar,\ystar)\cdot(y - \ystar)
\end{eqnarray*}
where $l_x$ and $l_y$ denote the partial derivatives of $l$ with
respect to $x$ and $y$, respectively.  Thus, we can
bound~\eqref{eq:1} by,
\begin{eqnarray}\label{eq:2}
F(q,Q) - F(\qstar,\Qstar) &\leq&
\int_{\bs\in\RR^M}\int_{\bu\in\RR^M}\int_{\by\in
\RR^N}\int_{\bH\in\cR_H}f(\bs)f(\by,\bH\given \bs, \bx = \bof(\bu,
\bs))\cdot\nonumber\nonumber\\&\cdot&
 [l_x(\qstar(\bu\given\bs),\Qstar(\bu\given\by,\bH))\cdot(q(\bu\given\bs)
- \qstar(\bu\given\bs)) +\nonumber\\&+&
l_y(\qstar(\bu\given\bs),\Qstar(\bu\given\by,\bH))\cdot(Q(\bu\given\by,\bH)
- \Qstar(\bu\given\by,\bH))]\dH\dy\du\ds
\end{eqnarray}
In the development below, we will show that this integral equals
zero.  This will then conclude the proof of the lemma.

To prove this, we will show that the two integrals below equal
zero.  For simplicity of notation, we let $q$ and $Q$ denote
$q(\bu\given\bs)$ and $Q(\bu\given\by,\bH)$, respectively.
\begin{eqnarray}
\hspace{-8mm}
\int_{\bs\in\RR^M}\int_{\bu\in\RR^M}\int_{\by\in
\RR^N}\int_{\bH\in\cR_H}f(\bs)f(\by,\bH\given \bs, \bx = \bof(\bu,
\bs))\cdot l_x(\qstar,\Qstar)\cdot (q-\qstar)\dHyus
&=& 0
\label{eq:3}\\
\hspace{-8mm}
\int_{\bs\in\RR^M}\int_{\bu\in\RR^M}\int_{\by\in
\RR^N}\int_{\bH\in\cR_H}f(\bs)f(\by,\bH\given \bs, \bx = \bof(\bu,
\bs))\cdot l_y(\qstar,\Qstar)\cdot (Q-\Qstar)\dHyus
&=& 0
\label{eq:4}
\end{eqnarray}
We first prove~\eqref{eq:3}.  Multiplying~\eqref{eq:Lagrange_q} by
$q - \qstar$, and using the fact that $l_x(x,y) = \log(y/x) -1$,
we get
\begin{eqnarray*}
& &\left[\int_{\by\in
\RR^N}\int_{\bH\in\cR_H}f(\bs)f(\by,\bH\given \bs, \bx = \bof(\bu,
\bs))l_x(\qstar,\Qstar)\dH \dy\right](q-\qstar) +\nonumber\\&
&\quad \left[f(\bs)<\Upsilon,\bof(\bu, \bs)\cdot\bof(\bu,
\bs)^T>\right](q-\qstar)+\left[f(\bs)<\Gamma,\bs\cdot\bof(\bu,
\bs)^T>\right](q-\qstar)+\alpha(\bs)(q-\qstar)
 = 0 \nonumber\\& &
\quad\quad\quad\quad\quad\quad\quad\quad\quad\quad\quad\quad\quad\quad\quad\quad\quad\quad\quad\quad\quad\quad\quad\quad\quad\quad\quad\quad\quad
\forall\bs\in\RR^M, \forall\bu\in\RR^M
\end{eqnarray*}
Integrating the above with respect to $\bu$ and $\bs$ would yield
zero.  We now focus on the integrals of the individual elements of
the above sum.  The first integral is equal to the left hand side
of~\eqref{eq:3}.  To prove this integral is zero, we will show
that the other integrals are zero. This will yield~\eqref{eq:3}.

We first integrate with respect to $\bu$ and then $\bs$.  The
order of integration matters, because the range of the integration
is unbounded, and some of the integrands are not non-negative and
not necessarily Lebesgue-integrable (i.e., the integral of their
absolute value may be infinite).
\begin{eqnarray*}
&&\int_{\bs\in\RR^M}\int_{\bu\in\RR^M}\left[f(\bs)<\Upsilon,\bof(\bu,
\bs)\cdot\bof(\bu, \bs)^T>\right](q-\qstar)\du\ds\\&&\quad=
<\Upsilon,\int_{\bs\in\RR^M}\int_{\bu\in\RR^M}f(\bs)\left[\bof(\bu,
\bs)\cdot\bof(\bu, \bs)^T\right]\cdot q\du\ds
\\ & & \quad -
\int_{\bs\in\RR^M}\int_{\bu\in\RR^M}f(\bs)\left[\bof(\bu,
\bs)\cdot\bof(\bu, \bs)^T\right]\cdot \qstar\du\ds>
\\&&\quad=<\Upsilon,\Sigma_X - \Sigma_X> = 0
\end{eqnarray*}
The equality before last results from~\eqref{eq:E_q} and from the
feasibility of the functions $q$ and $\qstar$.  In a similar way,
using~\eqref{eq:C_q}, we obtain that,
\begin{eqnarray*}
&&\int_{\bs\in\RR^M}\int_{\bu\in\RR^M}\left[f(\bs)<\Gamma,\bs\cdot\bof(\bu,
\bs)^T>\right](q-\qstar)\du\ds = 0
\end{eqnarray*}
Finally, we examine the last integral.
\begin{eqnarray*}
&&\int_{\bs\in\RR^M}\int_{\bu\in\RR^M}\alpha(\bs)(q-\qstar)\du\ds
= \int_{\bs\in\RR^M}\alpha(\bs)\left[\int_{\bu\in\RR^M}q\du
-\int_{\bu\in\RR^M}\qstar\du\right]\ds \\&&\quad =
\int_{\bs\in\RR^M}\alpha(\bs)\left[1 -1\right]\ds= 0
\end{eqnarray*}
The equality before last results from~\eqref{eq:q_valid}.  Thus,
we obtain~\eqref{eq:3}.

Similarly, relying on~\eqref{eq:Lagrange_Q}
and~\eqref{eq:Q_valid}, we obtain,
\begin{eqnarray}\label{eq:5}
\int_{\by\in
\RR^N}\int_{\bH\in\cR_H}\int_{\bu\in\RR^M}\int_{\bs\in\RR^M}f(\bs)f(\by,\bH\given
\bs, \bx = \bof(\bu, \bs))\cdot l_y(\qstar,\Qstar)\cdot
(Q-\Qstar)\ds\du\dH\dy= 0
\end{eqnarray}
The order of integration, unfortunately, is not that
of~\eqref{eq:4}.  To prove that we may change the order of
integration, we must prove that the integrand is
Lebesgue-integrable (Fubini's Theorem, see
e.g.~\cite{Billingsley}[Theorem~18.3]). To do this, we will prove
that
\begin{eqnarray}\label{eq:6}
\int_{\by\in
\RR^N}\int_{\bH\in\cR_H}\int_{\bu\in\RR^M}\int_{\bs\in\RR^M}f(\bs)f(\by,\bH\given
\bs, \bx = \bof(\bu, \bs))\cdot l_y(\qstar,\Qstar)\cdot
Q\ds\du\dH\dy < \infty
\end{eqnarray}
Since the integrand in the above is nonnegative, this would yield
that it is integrable.
Since $Q$ is arbitrary, the same would
apply if we replace it with $\Qstar$. The integrand
in~\eqref{eq:5}, which is not necessarily nonnegative, is thus also integrable because it is obtained by subtracting the integrand in~\eqref{eq:6} by the same expression, with $Q$ replaced by
$\Qstar$.

Using $l_y(x,y) = x/y$, we may rewrite the left hand side
of~\eqref{eq:6} as
\begin{eqnarray}
&&\int_{\by\in
\RR^N}\int_{\bH\in\cR_H}\int_{\bu\in\RR^M}\int_{\bs\in\RR^M}f(\bs)f(\by,\bH\given
\bs, \bx = \bof(\bu, \bs))\cdot \frac{\qstar}{\Qstar}\cdot
Q\ds\du\dH\dy \nonumber\\
&& \quad =\int_{\by\in \RR^N}\int_{\bH\in\cR_H}\int_{\bu\in\RR^M}
\frac{Q}{\Qstar}\cdot\left[\int_{\bs\in\RR^M}f(\bs)f(\by,\bH\given
\bs, \bx = \bof(\bu, \bs))\cdot\qstar \ds\right]\du\dH\dy
\label{eq:7}
\end{eqnarray}
The inside of the brackets is equal to $\qstar(\by,\bH,\bu)$,
defined to equal the marginal density of $\bY$, $\bH$ and $\bU$
where the distribution of $\bU$ given $\bS$ is determined by the
density $\qstar$.  Similarly defining $\qstar(\by,\bH)$, we obtain
by the conditions of Lemma~\ref{lemma:Lagrange}, that
$\qstar(\by,\bH,\bu) =
\qstar(\by,\bH)\cdot\Qstar(\bu\given\by,\bH)$.  Thus,~\eqref{eq:7}
becomes,
\begin{eqnarray*}
&&\int_{\by\in \RR^N}\int_{\bH\in\cR_H}\int_{\bu\in\RR^M}
\frac{Q}{\Qstar}\cdot\qstar(\by,\bH)\cdot \Qstar\du\dH\dy
=\int_{\by\in
\RR^N}\int_{\bH\in\cR_H}\qstar(\by,\bH)\int_{\bu\in\RR^M} Q
\du\dH\dy \\&&\quad=\int_{\by\in
\RR^N}\int_{\bH\in\cR_H}\qstar(\by,\bH)\cdot 1\dH\dy = 1 < \infty
\end{eqnarray*}
Thus, by the above discussion, the order of integration
in~\eqref{eq:5} can be changed, and we obtain~\eqref{eq:4}.
Coupled with~\eqref{eq:3}, this proves that the right hand side
of~\eqref{eq:2} is zero, concluding the proof of the lemma.
\QEDopen

\section{The Achievability of $F(q_G,Q_G)$}\label{apdx:Achievability_FqGQG}
The random variables $\bU, \bS, \bY, H$ that achieve the LAFP
capacity are continuous. In practice one can only realize the
Gelfand-Pinsker capacity of a set $\hat{\bU}, \hat{\bS},
\hat{\bY}, \hat{H}$ of discrete random variables. We now show that
$\bU, \bS, \bY, H$ can be quantized to a set $\hat{\bU},
\hat{\bS}, \hat{\bY}, \hat{H}$ of discrete random variables that
can approach the LAFP capacity arbitrarily close. The LAFP
capacity is given by $R_{\rm achievable} = F(q_G,Q_G)$ where
$F(q,Q)$ is defined by~\eqref{eq:F_q_Q_defined}.

We create a quantized version as follows. Let $\cB_n(c,d)$ denote
a cube in $\RR^n$ with center $c$ and size length $d$, i.e.,
$$
\cB_n(c,d) = \left\{ (x_1,x_2,\ldots,x_n) \: : \: c-d/2 < x_i \le
c+d/2,\: i=1,\ldots,n \right\}
$$
We define discrete random variables
$\hat{\bS},\hat{\bU},\hat{\bY},\hat{H}$ which are quantized
versions of ${\bS},{\bU},{\bY},{H}$, respectively, as follows.
Recall that $M$ and $N$ are the dimensions of $\bS$ and $\bY$,
respectively. The dimension of $H$ is thus $M\times N$.
Fix some $\epsilon>0$ sufficiently small, and
$\rho>0$ sufficiently large. Let $\bs_i$, $i=1,\ldots,N_s$ denote
all the points in $\RR^M$, such that $\bs_i \in
\cB_M(0,\rho)$ and such that all the coordinates of $\bs_i$ are
integer multiples of $\epsilon$. Similarly, let $\bu_j$,
$j=1,\ldots,N_u$, $\by_k$, $k=1,\ldots,N_y$ and $\bH_l$,
$l=1,\ldots,N_h$ denote all the points in $\RR^M$, $\RR^N$
and ${\cal R}_H$, such that $\bu_j \in \cB_M(0,\rho)$,
$\by_k \in \cB_N(0,\rho)$ and $\bH_l \in \cB_{MN}(0,\rho)$, and
such that all the coordinates of $\bu_j$, $\by_k$ and $\bH_l$ are
integer multiples of $\epsilon$.

We define by $\cS_i$, $i=0,1,\ldots,N_s$ the following regions,
$$
\cS_i = \left\{
          \begin{array}{ll}
            \RR^M \bigcap \cB_M(\bs_i,\epsilon), & \hbox{if $i=1,2,\ldots,N_s$;} \\
            \RR^M \setminus \left[{\bigcup_{i=1}^{N_s} \cB_M(\bs_i,\epsilon)}\right], & \hbox{if $i=0$.}
          \end{array}
        \right.
$$
Similarly we define
$$
\cU_j = \left\{
          \begin{array}{ll}
            \RR^M \bigcap \cB_M(\bu_j,\epsilon), & \hbox{if $j=1,2,\ldots,N_u$;} \\
            \RR^M \setminus \left[{\bigcup_{j=1}^{N_u} \cB_M(\bu_j,\epsilon)}\right], & \hbox{if $j=0$.}
          \end{array}
        \right.
$$
$$
\cY_k = \left\{
          \begin{array}{ll}
            \cB_N(\by_k,\epsilon), & \hbox{if $k=1,2,\ldots,N_y$;} \\
            \RR^N \setminus {\bigcup_{k=1}^{N_y} \cB_N(\by_k,\epsilon)}, & \hbox{if $k=0$.}
          \end{array}
        \right.
$$
and
$$
\cH_l = \left\{
          \begin{array}{ll}
            \cR_H \bigcap \cB_{MN}(\bH_l,\epsilon), & \hbox{if $l=1,2,\ldots,N_h$;} \\
            \cR_H \setminus{\bigcup_{l=1}^{N_h} \cB_{M N}(\bH_l,\epsilon)}, & \hbox{if $l=0$.}
          \end{array}
        \right.
$$
The quantized random variable $\hat{\bS}$ is defined as follows:
$\hat{\bS} = i$ if $\bS \in \cS_i$. The quantized random variables
$\hat{\bU}$, $\hat{\bY}$ and $\hat{H}$ are defined similarly. The
joint probability of $\hat{\bS},\hat{\bU},\hat{\bY},\hat{H}$ is,
\baa \lefteqn{ P \left( \hat{\bS}=i, \hat{\bU}=j, \hat{\bY}=k,
\hat{H}=l \right) = }
\\ & &
\int_{\bs \in \cS_i} \int_{\bu \in \cU_j} \int_{\by \in \cY_k}
\int_{\bH \in \cH_l} f(\bs) f(\by, \bH \given \bs,\bx =
{\bf}(\bs,\bu)) q_G(\bu \given \bs) \: d \bH \: d{\by} \: d{\bu}
\: d{\bs} \eaa

The Gelfand-Pinsker achievable rate corresponding to the quantized
random variables is, \bre \hat{R} = \sum_{i,j,k,l} P \left(
\hat{\bS}=i, \hat{\bU}=j, \hat{\bY}=k, \hat{H}=l \right) \log
\frac{P(\hat{\bU}=j \given \hat{\bY}=k, \hat{H}=l)} {P(\hat{\bU}=j
\given \hat{\bS}=i)} \label{eq:hR} \ere

We claim that $\hat{R} = R_{\rm achievable} +
o_{\epsilon,\rho}(1)$ where $o_{\epsilon,\rho}(1)$ is a term that
approaches $0$ as $\epsilon \rightarrow 0$ and $\rho\rightarrow
\infty$.

To see this, first note that when $\bs \in \cS_0$ or $\bu \in
\cU_0$ or $\by \in \cY_0$ or $\bH \in \cH_0$, the contribution to
$F(q_G,Q_G)$ in~(\ref{eq:F_q_Q_defined}) approaches $0$ as
$\rho\rightarrow \infty$. In addition, $\log \frac{Q_G(\bu \given
\by, \bH)}{q_G(\bu \given \bs)}$ is uniformly continuous in the
region $\bs \in \overline{\cS_0}$, $\bu \in \overline{\cU_0}$,
$\by \in \overline{\cY_0}$, $\bH \in \overline{\cH_0}$. Hence,
$$
F(q_G,Q_G) = \sum_{i\ne 0,j\ne 0,k\ne 0,l\ne 0} P \left(
\hat{\bS}=i, \hat{\bU}=j, \hat{\bY}=k, \hat{H}=l \right) \log
\frac{Q_G(\bu_j \given \by_k, \bH_l)}{q_G(\bu_j \given \bs_i)} +
o_{\epsilon,\rho}(1)
$$
In addition, by the uniform continuity of the Gaussian
distribution in the region $\bs \in \overline{\cS_0}$, $\bu \in
\overline{\cU_0}$, $\by \in \overline{\cY_0}$, $\bH \in
\overline{\cH_0}$,
$$
\frac{P(\hat{\bU}=j \given \hat{\bY}=k, \hat{H}=l)}{Q_G(\bu_j
\given \by_k, \bH_l)} = 1 + o_{\epsilon,\rho}(1)
$$
and
$$
\frac{P(\hat{\bU}=j \given \hat{\bs}=i)}{q_G(\bu_j \given \bs_i)}
= 1 + o_{\epsilon,\rho}(1)
$$

Finally by arguments similar to those indicated above, the
contribution of terms with $i=0$ or $j=0$ or $k=0$ or $l=0$
in~(\ref{eq:hR}) is negligible.

Hence we obtained the desired claim that $\hat{R} = R_{\rm
achievable} + o_{\epsilon,\rho}(1)$.
\section{Proof of Theorem~\ref{theorem:Independent_Maximizes}}
\label{apdx:Proof_Independent_Maximizes}  Our approach is the
following. We begin with an assignment of variables for the LAFP
achievable region.  This means a set of variables
$\bX_1$,...,$\bX_L$ that are not necessarily independent.  A set
of matrices $\bF_1,...,\bF_L$ and a set of auxiliary random
variables $\bU_l = \bF_l\bS_l + \bX_l$ where $\bS_l = \Sigma_{i <
l}\bX_i$.  Recall that in our current context, $\bX =
\bX_1+...+\bX_L$ denotes the transmitted symbol of the MIMO-BC
channel, while $\bX_l$ denotes the transmitted signal to user $l$,
equivalent to $\bX$ as in Sec.~\ref{sec:Fading_Lin_Strat}.

We will construct an alternative set of independent random
variables $\hbX_1,...\hbX_L$ and $\hbF_1,...,\hbF_L$ such that the
transmitted signal $\hbX
\defined \hbX_1+...+\hbX_L = \bX_1+...+\bX_L
= \bX$. Thus, the distribution of the actual transmitted signal is
unchanged and satisfies the power constraint.  Furthermore, we
show that for similarly defined $\hbU_l= \hbF_l\hbS_l + \hbX_l$
and $\hbS_l = \Sigma_{i < l}\hbX_i$, the achievable rates satisfy
$\hR_l \geq R_l$, where
\begin{eqnarray*}
\hR_l \defined I(\hbU_l;\bY_l,\bH_l) - I(\hbU_l;\hbS_l),\quad\quad
R_l = I(\bU_l;\bY_l,\bH_l) - I(\bU_l;\bS_l)
\end{eqnarray*}
\subsection{Definition of $\hbX_1,...,\hbX_L$}
For each $l = 1,...,L$, using Gram-Schmidt orthogonalization,
$\bX_l$ can be written as $\bX_l = \Gamma_l\bS_l + \bX'_l$ where
$\Gamma_l$ is a matrix and where $\bS_l$ and $\bX'_l$ are
uncorrelated. Therefore, since we have assumed, in our definition
of the LAFP region in Sec.~\ref{subsec:LAFP_Def}, that all
variables are jointly Gaussian, they are independent. With this
definition,
\begin{eqnarray*}
\bX &=& \bS_L + \bX_L = (\bI + \Gamma_L)\bS_L + \bX'_L
\\&=& (\bI+\Gamma_L)\left[\bS_{L-1} +
\bX_{L-1}\right] + \bX'_L=
(\bI+\Gamma_L)\left[(\bI+\Gamma_{L-1})\bS_{L-1} +
\bX'_{L-1}\right] +
\bX'_L \\&&...\\
&=& (\bI+\Gamma_L)\cdot...\cdot(\bI+\Gamma_2)\bX'_1 +
(\bI+\Gamma_L)\cdot...\cdot(\bI+\Gamma_3)\bX'_2 + ... +
(\bI+\Gamma_L)\bX'_{L-1}  +\bX'_L
\end{eqnarray*}
We thus define $\hbX_l = \bG_l\bX'_l$ where $\bG_l
=(\bI+\Gamma_L)\cdot...\cdot(\bI+\Gamma_{l+1})$ $l = 1,...,L-1$,
$\bG_L = \bI$.  By construction, $\sum_{l=1}^L\hbX_l = \bX =
\sum_{l=1}^L\bX_l$, as desired.

The following lemma summarized some properties of our random
variables.
\begin{lemma} \label{lemma:Properties_HBX_HSX}For all $l = 1,...,L$,
\begin{enumerate}
\item \label{enum:1} $\hbX_l$ is independent of $\bX_1,...,\bX_{l-1}$.
\item \label{enum:2} $\hbX_l$ is independent of $\bS_1,...,\bS_l$.
\item \label{enum:3} $\hbX_l$ is independent of $\hbX_1,...,\hbX_{l-1}$.
\item \label{enum:4} $\hbS_l = G_{l-1}\bS_l $
\end{enumerate}
\end{lemma}
\beginproof To prove property~\ref{enum:1}, observe that the following
Markov relations hold: $\bX_1,...,\bX_{l-1} \longleftrightarrow
\bS_l \longleftrightarrow \bS_l,\bX_l \longleftrightarrow \hbX_l$.
$\hbX_l$, by construction, is independent of $\bS_l$.  It is thus
straightforward to verify, using this Markov relation, that it is
also independent of $\bX_1,...,\bX_{l-1}$.  To obtain
properties~\ref{enum:2} and~\ref{enum:3}, observe that
$\bS_1,...,\bS_l$ and $\hbX_1,...,\hbX_{l-1}$ are functions of
$\bX_1,...,\bX_{l-1}$ and thus are independent of $\hbX_l$.

The last property is easily obtained by induction.  For $l = 1$,
\begin{eqnarray*}
\hbS_1 = \sum_{i < 1} \hbX_i = \bzr = \sum_{i < 1} \bX_i = \bS_1
\end{eqnarray*}
The rest is obtained by the following induction:
\begin{eqnarray*}
\hbS_{l+1} &=& \hbS_l + \hbX_l = G_{l-1}\bS_l + G_l\bX'_l = G_l(I
+ \Gamma_l)\bS_l + G_l\bX'_l = G_l\left[(I + \Gamma_l)\bS_l +
\bX'_l \right] \\&=& G_l\left[\bS_l + \Gamma_l\bS_l + \bX'_l \right]
= G_l\left[\bS_l + \bX_l\right] = G_l\bS_{l+1}
\end{eqnarray*}
 \QEDopen

\subsection{Definition of $\hbF_1,...,\hbF_L$}
We have not yet defined $\hbF_l$. To do so, we first consider
$\bG_l\cdot \bU_l$. By the definition of $\bU_l$
\begin{eqnarray} \label{eq:16}
\bG_l\cdot \bU_l = \bG_l[\bF_l \bS_l + \bX_l] = \bG_l[(\bF_l +
\Gamma_l)\bS_l + \bX'_l] = \bG_l(\bF_l + \Gamma_l)\bS_l + \hbX_l
\end{eqnarray}
where the last inequality was obtained by the definition of
$\hbX_l$, above.  Using Gram-Schmidt orthogonalization, $\bS_l$
can be written as $\bS_l = \bB_l\cdot \hbS_l + \bD_l$ where
$\bB_l$ is a matrix and $\bD_l$ is uncorrelated with $\hbS_l$.
Since the variables are jointly Gaussian, $\bD_l$ is also
independent of $\hbS_l$. We proceed
\begin{eqnarray}
\bG_l\cdot \bU_l &=& \bG_l(\bF_l + \Gamma_l)[\bB_l\hbS_l + \bD_l] + \hbX_l \nonumber\\
&=& \bG_l(\bF_l + \Gamma_l)\bB_l\hbS_l  + \hbX_l + \bG_l(\bF_l +
\Gamma_l)\bD_l \label{eq:17}
\end{eqnarray}
We define $\hbF_l = \bG_l(\bF_l + \Gamma_l)\bB_l$.
\subsection{Proof of $\hR_l > R_l$}
Recall that $\hbU_l = \hbF_l\hbS_l + \hbX_l$.  To prove $\hR_l >
R_l$, we first define an intermediate auxiliary variable $\tbU_l =
\bG_l\cdot \bU_l$. Since $\tbU_l$ is a function of $\bU_l$, we
have
\begin{eqnarray*}
R_l &=& I(\tbU_l, \bU_l; \bY_l, H_l) - I(\tbU_l, \bU_l; \bS_l)
\\&=& H(\tbU_l, \bU_l \given \bS_l) - H(\tbU_l, \bU_l\given
\bY_l, H_l)\\
&=& H(\tbU_l \given \bS_l) - H(\tbU_l\given \bY_l, H_l) + H(\bU_l
\given \tbU_l, \bS_l) - H(\bU_l \given \tbU_l, \bY_l,
H_l)\\
&=& \left[H(\tbU_l \given \bS_l) - H(\tbU_l\given \bY_l,
H_l)\right] + \left[I(\bU_l; \tbU_l, \bY_l, H_l) - I(\bU_l ;
\tbU_l, \bS_l)\right]
\end{eqnarray*}
We now wish to show that the contents of the second brackets are
non-positive. For this purpose, we will show that the following
Markov relations hold: $\bU_l \longleftrightarrow \tbU_l, \bS_l
\longleftrightarrow \tbU_l, \hbX_l, \hbS_l \longleftrightarrow
\tbU_l, \bX \longleftrightarrow \tbU_l, \bY_l, H_l$.  The desired
result will then follow from the first and last Markov relations,
using the data processing inequality.

The second relation (first Markov triple) follows from the fact
that $\hbX_l$ and $\hbS_l$ may be determined from $\tbU_l$ and
$\bS_l$ by means of deterministic functions: $\hbX_l$
through~\eqref{eq:16}, and $\hbS_l$, by
Lemma~\ref{lemma:Properties_HBX_HSX} satisfies $\hbS_l = \bG_{l-1}
\bS_l$. For the third relation, observe that $\bX = \hbS_l +
\hbX_l + \sum_{i
> l} \hbX_i$.  By the above definition all $\{\hbX_i\}_{i > l}$,
are independent of $\bU_l, \tbU_l, \hbS_l, \bS_l$ and $\hbX_l$.
Therefore this Markov relation holds. The last Markov relation is
straightforward.

We thus have,
\begin{eqnarray}\label{eq:18}
R_l \leq H(\tbU_l \given \bS_l) - H(\tbU_l\given \bY_l, H_l)
\end{eqnarray}
Examining the first element of the above difference, we obtain:
\begin{eqnarray}
H(\tbU_l \given \bS_l) &=& H(\bG_l(\bF_l + \Gamma_l)\bS_l + \hbX_l
\given \bS_l) = H(\hbX_l \given \bS_l) = H(\hbX_l) = H(\hbX_l
\given \hbS_l) = H(\hbF_l\hbS_l + \hbX_l \given \hbS_l) \nonumber
\\&=& H(\hbU_l \given \hbS_l) \label{eq:19}
\end{eqnarray}
where the first equality follows from the definition of $\tbU_l$
and from~\eqref{eq:16}.  The third equality follows from the
independence of $\hbX_l$ and $\bS_l$ and the fourth from the
independence of $\hbX_l$ and $\hbS_l$.

Examining the second element of~\eqref{eq:18}, we have
\begin{eqnarray}\label{eq:20}
H(\tbU_l\given \bY_l, H_l) &=& H(\hbU_l + \bG_l(\bF_l +
\Gamma_l)\bD_l \given \bY_l, H_l) \geq H(\hbU_l + \bG_l(\bF_l +
\Gamma_l)\bD_l \given \bY_l, H_l, \bD_l) \nonumber\\&=&
H(\hbU_l\given \bY_l, H_l, \bD_l) = H(\hbU_l \given \bY_l, H_l)
\end{eqnarray}
The first equality follows from~\eqref{eq:17} and the definitions
of $\tbU_l$, $\hbU_l$ and $\hbF_l$.  The inequality results from
the fact that conditioning cannot increase the entropy.  To prove
the last equality, we wish to show that $\bD_l$ and $\hbU_l$ are
independent, given $\bY_l$ and $H_l$.

$\hbU_l$ is a function of $\hbS_l$ and $\hbX_l$. Therefore, it
suffices to show that $\bD_l$ is independent of these two random
variables, given $\bY_l$ and $H_l$. $\bD_l$ is independent of
$\hbS_l$ by construction.  In addition, $\hbX_l$ is independent of
$\bS_l,\hbS_l$ and $\bD_l$, because $\bD_l$ is a function of
$\bS_l$ and of $\hbS_l$, where $\hbS_l = \bG_{l-1}\bS_l$ (by
Lemma~\ref{lemma:Properties_HBX_HSX}), and $\hbX_l$ is independent
of $\bS_l$ (again, by Lemma~\ref{lemma:Properties_HBX_HSX}).
Therefore, $\bD_l$ is independent of $\hbS_l$ and $\hbX_l$.  To
show that the independence is maintained even when we condition by
$\bY_l$ and $H_l$, we prove the following Markov chain relation $\bD_l
\longleftrightarrow \hbS_l, \hbX_l \longleftrightarrow
\hbS_l,\hbX_l,...\hbX_L \longleftrightarrow \bX
\longleftrightarrow H_l, \bY_l$.  The second relation (first
Markov triple) holds because the random variables
$\hbX_{l+1},...,\hbX_L$ are independent of $\hbS_l$ and of
$\hbX_l$ by Lemma~\ref{lemma:Properties_HBX_HSX}, and of $\bD_l$,
by virtue of it being a function of $\hbS_l$ and $\bS_l$.  The
third relation holds because $\bX = \hbS_l + \hbX_l + ... +
\hbX_L$.  The fourth relation holds because $\bY_l = H_l\bX +
\bZ_l$ and $H_l$ and $\bZ_l$ are independent of the other random
variables $\bD_l, \hbS_l, \hbX_l,...\hbX_L, \bX$.

Combining \eqref{eq:18},~\eqref{eq:19} and~\eqref{eq:20} we
obtain,
\begin{eqnarray*}
R_l \leq H(\hbU_l \given \hbS_l) - H(\hbU_l\given \bY_l, H_l) =
 I(\hbU_l; \bY_l, H_l) - I(\hbU_l ; \hbS_l) = \hR_l
\end{eqnarray*}
This completes the proof. \QEDopen

\section{Computing the LAFP Achievable Region when the Number of Users is Greater than Two} \label{apdx:LAFP_L_GT_2}
In Sec.~\ref{sec:Comparison Dirty Paper} we considered the
computation of the LAFP achievable region over a fading MIMO-BC
channel where the number of users is two.  In this appendix we
briefly consider the case of more than two users. To obtain the
LAFP achievable region, we could again (as in
Sec.~\ref{sec:Comparison Dirty Paper}) apply a grid search to
obtain $\{\Sigma_X^{(l)}\}_{l=1}^{L}$. A straightforward approach
would be to compute, for each choice of such matrices, the
achievable rates for each of the individual users by selecting the
matrices $\bF$, for each user (except for the first who does not
have an associated $\bF$ matrix) so as to
maximize~(\ref{eq:Gauss_LA_Achievable}). However, the
computational complexity of such an approach would grow
exponentially with the number of users.

The following observation can be used to reduce the number of
computations.  The achievable rate for user $l$ is a function of
$\Sigma_X^{(l)}$ (the covariance matrix of its transmitted signal
$\bX_l$), of $\Sigma_S^{(l)}\defined \sum_{i<l} \Sigma_X^{(i)}$
(the covariance matrix of the interference $\bS_l =
\sum_{i<l}\bX_i$) and $\Sigma_Z^{(l)}\defined \sum_{i>l}
H \Sigma_X^{(i)} H^T + \bI$
(the covariance matrix of the
effective noise $\bZ_l = H \sum_{i>l}\bX_i + \bZ$).  Thus, the
achievable rate for user $l$ needs to be computed only once for each of the
possible choices of $\Sigma_S^{(l)}$, $\Sigma_X^{(l)}$ and
$\Sigma_Z^{(l)}$, and not for each choice of
$\{\Sigma_X^{(i)}\}_{i=1}^{L}$.  A dynamic-programming algorithm
that relies on this observation can dramatically reduce the number
of computations.
This approach is useful when the number of transmit antennas and the number of receive antennas of each user is small (the number of users can be large). Otherwise we can resort to suboptimal methods for computing the transmit covariances  $\{\Sigma_X^{(l)}\}_{l=1}^{L}$ (and the $\bF$ matrices), e.g. using gradient descent or alternate maximization that maximizes the sum rate with respect to two $\Sigma_X^{(l)}$-s at a time, while fixing the other $\Sigma_X^{(l)}$-s.
\section{Computing a Cooperative Upper-Bound in our Setting} \label{apdx:Sato}
Sato's upper bound~\cite{Sato} on the sum rate capacity (the
maximum achievable sum-rate) of a broadcast channel relies on two
observations:
\begin{enumerate}
\item A fundamental assumption in the broadcast channel model is
that the users are not able to cooperate in their decoding.
Consider a virtual channel where the users are allowed to
cooperate.  The sum capacity in this channel is clearly an upper
bound on the sum rate capacity of the true channel.  Such a
cooperative model is equivalent to transmission to a single
virtual user, to whom all the outputs of the broadcast channel
users are made available.
\item The capacity region of a broadcast channel depends not on
the joint distribution $\Pr(\bY_1,H_1,...,\bY_L,H_L\given \bX)$
but on the marginal distributions $\Pr(\bY_1,H_1\given \bX
),...,\Pr(\bY_L,H_L\given \bX)$ alone.  Thus, we may alter our
model by introducing correlation between the noise signals and
channel matrices of different users.  As long as the marginal
statistics of the individual channels to each of the users stay
the same, the resulting broadcast channel's capacity region will
remain unchanged. However, introducing correlations {\it could}
alter (and tighten) the above-mentioned cooperative upper bound.
\end{enumerate}
Note that with any valid choice of correlation that we choose to
introduce, the maximum cooperative sum-rate produces an upper
bound on the broadcast channel's sum-rate capacity.  We refer to
such an upper bound as a {\it cooperative} upper bound.  The Sato
upper bound is the tightest such bound.

Consider the channel to the virtual single user corresponding to
the fading MIMO-BC example of Sec.~\ref{sec:Comparison Dirty
Paper}. This user will observe a virtual channel matrix and a
virtual noise defined as,
\begin{eqnarray*}
H = \left[\begin{array}{c}
    H^{(1)}\\
    H^{(2)}
\end{array}\right]\quad \textrm{and}\quad \bZ =\left[\begin{array}{c}
    Z^{(1)}\\
    Z^{(2)}
\end{array}\right]
\end{eqnarray*}
Our above discussion implies that we may freely introduce
correlations as long as we do not alter the statistics of the
channel observed by each of the individual users.  We may thus
introduce a correlation between the two noise signals $Z^{(1)}$
and $Z^{(2)}$, following the examples of~\cite{Caire_Shamai_MIMO}
and~\cite{Goldsmith_MIMO_BC}.  We may also introduce correlation
between the two channel matrices $H^{(1)}$ and $H^{(2)}$.
Furthermore, we may introduce correlation between the channel
matrix of one user and the noise of the other.

The possible values for $H$ are,
\begin{eqnarray*}
H \in \left\{ \bH_1 = \left[\begin{array}{cc}
  1 &  0.4 \\
  0.4 & 1
\end{array}\right], \bH_2 = \left[\begin{array}{cc}
  1 &  0.4 \\
  3 & 1
\end{array}\right],\bH_3 = \left[\begin{array}{cc}
  1 &  3     \\
  0.4 & 1
\end{array}\right],\bH_4 = \left[\begin{array}{cc}
  1 &  3 \\
  3 & 1
\end{array}\right] \right\}
\end{eqnarray*}
Let $p(H)$ denote the probability assignment to each of the above
matrices.  To preserve the marginal statistics of the channel to
each of the individual users, we require that $p(H)$ satisfy the
following constraints,
\begin{eqnarray*}
p(\bH_1) + p(\bH_2) = \frac{1}{2}\\
p(\bH_1) + p(\bH_3) = \frac{1}{2}
\end{eqnarray*}
Furthermore, for $p(H)$ to be a valid probability assignment, it
must satisfy,
\begin{eqnarray*}
p(\bH_1) + p(\bH_2) + p(\bH_3) + p(\bH_4) = 1
\end{eqnarray*}
The constraints imply that $p(H)$ is completely described by
$\alpha \defined p(\bH_1)$.  That is, for any $\alpha \in [0,
1/2]$, we have
\begin{eqnarray*}
p(\bH_1) = p(\bH_4) = \alpha,\quad p(\bH_2) =p(\bH_3) =
\frac{1}{2}-\alpha
\end{eqnarray*}
One way to introduce correlation between the various noise
elements is to follow the approach of~\cite{Caire_Shamai_MIMO}.
That is, introduce a correlation coefficient $\rho \in (-1,1)$ and
consider a virtual noise $\bZ$ whose covariance matrix is,
\begin{eqnarray*}
\Cov(\bZ) = \left[\begin{array}{cc}
  1 &  \rho \\
  \rho & 1
\end{array}\right]
\end{eqnarray*}
However, a more general approach would introduce correlation
between the virtual noise and the above virtual channel matrix in
the following way:  We will consider {\it four} correlation
coefficients $\rho_1,\rho_2,\rho_3,\rho_4$ such that,
\begin{eqnarray*}
\Cov(\bZ\given H = \bH_i) = \left[\begin{array}{cc}
  1 &  \rho_i \\
  \rho_i & 1
\end{array}\right]\quad i = 1,...,4
\end{eqnarray*}
The channel noise observed by each of the users remains
distributed as $\cN(0,1)$.  Furthermore, each of the individual
realizations of $Z^{(1)}$ and $Z^{(2)}$ remains independent of the
respective channel matrices $H^{(1)}$ and $H^{(2)}$.  Thus, the
marginal statistics of the channels to each of the individual
users remain unchanged, as desired.

The capacity of the channel to the virtual user is now obtain by
taking the maximum of,
\begin{eqnarray*}
I(\bX; \bY,H) = I(\bX; H) + I(\bX;\bY\given H)= I(\bX;\bY\given H)
=  \sum_{i=1}^4 p(\bH_i) I(\bX;\bY\given H = \bH_i)
\end{eqnarray*}
The first equality is obtained by the chain rule for mutual
information, and the second by the independence of $\bX$ and $H$.
The distribution that maximizes the above is clearly Gaussian.
Thus,
\begin{eqnarray}\label{eq:C_Coop}
C = \max_{\Sigma_X} \sum_{i=1}^4 p(\bH_i)
\frac{1}{2}\log\det\left(\bI + \Lambda_{i}^{-1/2}\bH_i \Sigma_X
\bH_i^T\Lambda_{i}^{-1/2}\right)
\end{eqnarray}
where $\Lambda_i \defined \Cov(\bZ \given H = \bH_i)$.

We may now numerically obtain a cooperative upper bound in the
following way.  We consider all choices of
$\alpha,\rho_1,...,\rho_4$ along a fine grid.  For each such
choice, we evaluate~\eqref{eq:C_Coop} by applying semidefinite
programming to determine the $\Sigma_X$ that achieves the maximum.
Each choice of $\alpha,\rho_1,...,\rho_4$ produces a cooperative
bound. We conclude by selecting the lowest (tightest)
bound\footnote{Note that the bound obtained in this way is not
necessarily the true Sato upper bound (i.e., the tightest possible
cooperative bound), because we have not proven that our approach
exhausts all the possible ways of introducing valid correlations
between the various signals.}.

In our numerical results (as presented in Sec.~\ref{sec:Comparison
Dirty Paper}), the tightest bound was obtained by setting $\alpha
= 0$ and $\rho_1=\rho_2=\rho_3=\rho_4 = 0.3$.  Thus, the tightest
bound was obtained with a limited exploitation of the available
degrees of freedom in the above approach.

\section*{Acknowledgements}
We would like to thank Marc Teboulle for helpful discussions, and
the anonymous reviewers for their comments, that helped improve
the presentation of the paper.
\begin{small}

\end{small}

\clearpage
%
%
%
%
%
%

\clearpage \listoftables (None)
 \listoffigures
\end{document}